\documentclass[peerreview,onecolumn]{IEEEtran}
\usepackage[T1]{fontenc}

\usepackage[utf8]{inputenc}
\usepackage{amsmath}
\usepackage{mathtools}
\usepackage{graphicx}
\usepackage{amsfonts}
\usepackage{color}
\usepackage{microtype} 
\usepackage{subcaption} 
\usepackage{array}
\usepackage{tabularx}
\usepackage{booktabs} 
\usepackage{cryptocode}

\usepackage{enumitem}  
\usepackage{verbatim} 

\usepackage{xparse} 
\usepackage{xspace}
\usepackage{braket}
\usepackage{physics}
\usepackage{bm}

\usepackage{float}
\usepackage{mdframed}

\makeindex
\usepackage{amsthm}
\usepackage{thmtools}
\usepackage{thm-restate}
\usepackage{tikz}
\usetikzlibrary{positioning, calc, shapes.geometric, backgrounds,fit}
\usetikzlibrary{decorations.pathreplacing}

\usepackage{titlesec}

\titleformat*{\paragraph}{\bfseries}

\usepackage{hyperref} 
\usepackage[capitalize,noabbrev]{cleveref} 

\crefname{enumi}{}{} 
\hypersetup{
	colorlinks = true,
	linkcolor =blue,
	citecolor=blue, 
	urlcolor=blue 
}

\newlength\colsep   
\newlength\offsetX  
\newlength\offsetY  

\newtheorem{lemma}{Lemma}[section]
\newtheorem{corollary}{Corollary}[section]

\newtheorem{remark}{Remark}[section]

\theoremstyle{definition} 
\newtheorem{definition}{Definition}[section]

\newtheorem{prot}{Protocol}

\newcommand{\realistic}{practical}

\newcommand{\keyspace}{\mathcal{K}}

\newcommand{\defvar}{\coloneqq} 

\newcommand{\dop}[1]{\operatorname{S}_{#1}} 

\newcommand{\epssecure}{\varepsilon^{\mathrm{secure}}}

\newcommand{\CAs}{\bm{{C_{A \rightarrow E}}} }
\newcommand{\CBs}{\bm{{C_{B \rightarrow E}}} }
\newcommand{\CAr}{\bm{{C_{E \rightarrow A}}} }
\newcommand{\CBr}{\bm{{C_{E \rightarrow B}}} }

\newcommand{\Cauth}{\bm{C_\mathrm{auth}}}
\newcommand{\CAsauth}{\bm{{C^\mathrm{auth}_{A \rightarrow E}}} }
\newcommand{\CBsauth}{ \bm{{C^\mathrm{auth}_{B \rightarrow E}}} }
\newcommand{\CArauth}{\bm{{C^\mathrm{auth}_{E \rightarrow A}}} }
\newcommand{\CBrauth}{\bm{ {C^\mathrm{auth}_{E \rightarrow B}}} }

\newcommand{\complement}{\mathrm{C}} 
\newcommand{\transcript}{\mathcal{T}}

\newcommand{\accept}{\monofont{accept}\xspace}
\newcommand{\abort}{\monofont{abort}\xspace}
\newcommand{\authabort}{\monofont{auth-abort}\xspace}

\newcommand{\tracedist}[1]{\frac{1}{2} \norm{#1}_1}

\newcommand{\monofont}[1]{\texttt{\textbf{#1}}}

\renewcommand{\selectlanguage}[1]{} 

\newcommand{\CPTP}{\operatorname{CPTP}} 

\newcommand{\authmap}
{\mathcal{E}^\mathrm{repl}_\mathrm{auth}}

\newcommand{\authcommmap}{\mathcal{E}_\mathrm{comm}}
\newcommand{\authupdatemap}{\mathcal{E}_\mathrm{update}}
\newcommand{\idealmap}{\mathcal{R}_\mathrm{ideal}}

\newcommand{\delauthcommmap}{\mathcal{E}^\mathrm{del}_\mathrm{comm}}
\newcommand{\delauthupdatemap}{\mathcal{E}^\mathrm{del}_\mathrm{update}}

\newcommand{\Esecdef}{\bm{E}}

\newcommand{\EfinalQKD}{\bm{E_\mathrm{fin}}}
 \newcommand{\CfinalQKD}{\bm{C_\mathrm{fin}}}  
 \newcommand{\Efinal}{\bm{E^\prime_{\mathrm{fin}}}} 
\newcommand{\Cfinal}{\bm{C^\prime_{\mathrm{fin}}}} 
\newcommand{\Ccorr}{\widetilde{C}}

\newcommand{\OlAlB}{\Omega_{l_A,l_B}}

\newcommand{\OlAlBprime}{\Omega_{l^\prime_A,l^\prime_B}}
\newcommand{\Onice}{\Omega_{\mathrm{auth\text{-}hon}}}
\newcommand{\Onicedel}{\Omega_{\mathrm{dauth\text{-}hon}}}

\newcommand{\QKDprotocol}{\mathcal{P}_\mathrm{QKD}}
\newcommand{\coreQKDprotocol}{\widetilde{\mathcal{P}}_\mathrm{QKD}}
\newcommand{\APPprotocol}{\mathcal{P}_\mathrm{APP}}
\newcommand{\delayedAPPprotocol}{\mathcal{P}^\mathrm{del}_\mathrm{APP}}
\newcommand{\delayedQKDprotocol}{\mathcal{P}^\mathrm{del}_\mathrm{QKD}}
\newcommand{\worldreal}{\mathcal{W}^\mathrm{real}_\mathrm{auth}}
\newcommand{\worlddelreal}{\mathcal{W}^\mathrm{real}_\mathrm{del-auth}}
\newcommand{\worldhonest}{\mathcal{W}^\mathrm{hon}_\mathrm{auth}}
\newcommand{\worldvirtual}{\mathcal{W}^\mathrm{virt}}
\newcommand{\projkeymap}{\Pi^{l^\prime_A,l^\prime_B}_{K_A K_B} }

\newcommand{\finalacceptabortevent}[2]{\Omega^\mathrm{fin-dec}_{(#1,#2)}}

\usepackage{tikz}

\usepackage{tikz-cd}

\usetikzlibrary{positioning,arrows.meta,shapes.geometric}

\tikzset{
  pics/detector/.style args={#1,#2,#3}{
    code={
      \draw[line width=0.5mm] (0,-1) -- (0,1);
      \draw[line width=0.5mm] (0,1) -- ++(0.1,0);
      \draw[line width=0.5mm] (0,-1) -- ++(0.1,0);
      \draw[line width=0.5mm] (0.1,1) arc (90:-90:1); 
      \node[#3, align=center] (#1) at (0.5,0) {#2};
    }
  }
}

\tikzset{
  pics/beamsplitter/.style args={#1,#2,#3}{
    code={
    \coordinate (A) at (0.3,0.3);
    \coordinate (B) at (-0.3,-0.3);
    \coordinate (#1) at (0,0);
    
    \draw[line width=0.3mm] (A) -- (B); 
    \node[align=center,font=\fontsize{7}{7}\selectfont] at #3 {#2};
    }
  }
}

\tikzset{
  pics/pulse/.style args = {#1}{
    code ={
    \draw[#1] plot[smooth,tension=1] coordinates {(0,0) (0.3,0.2) (0.5,1) (0.7,0.2) (1,0)};
    }
  }
}

\tikzstyle{process} = [rectangle, line width=0.3mm, minimum width=1cm, minimum height=1cm, text centered, text width=1cm, draw=black]

\begin{document}
	\title{Authentication in Security Proofs for Quantum Key Distribution}

\author{Devashish Tupkary, Shlok Nahar, Ernest Y.-Z. Tan%
\thanks{D. Tupkary and S. Nahar are with the Institute for Quantum Computing and the Department of Physics and Astronomy, University of Waterloo, Waterloo, ON, Canada (e-mail: djtupkary@uwaterloo.ca; sanahar@uwaterloo.ca).}%
\thanks{E. Y.-Z. Tan is with the Department of Physics, National University of Singapore, Singapore 117542, and was also with the Institute for Quantum Computing and the Department of Physics and Astronomy, University of Waterloo, Waterloo, ON, Canada (e-mail: phytyze@nus.edu.sg).}%
}

\maketitle

\begin{abstract}
Quantum Key Distribution (QKD) protocols rely on authenticated classical communication. Typical QKD security proofs are carried out in an idealized setting where authentication is assumed to behave honestly: it never aborts, and all classical messages are delivered faithfully with their original timing preserved.
Authenticated channels that can be constructed in practice have different properties. Most critically, such channels may abort asymmetrically, such that only the receiving party may detect an authentication failure while the sending party remains unaware. Furthermore,  an adversary may delay, reorder, or block classical messages.
This discrepancy means that existing security proofs under honest authentication do not directly establish security in the {\realistic} authentication setting. We address this gap using an existing security definition from the literature that permits asymmetric aborts while retaining secrecy of any key produced and agreement of the keys whenever both parties accept. Our main result is a reduction theorem showing that, under mild and easily satisfied protocol conditions, any QKD protocol proven secure under the honest authentication setting, for the standard security definition with symmetric aborts, remains secure under a {\realistic} authentication setting, for the weaker security definition that allows for asymmetric aborts. This result
allows all existing QKD proofs to be retroactively lifted to the practical authentication setting (with asymmetric aborts) with
a minor protocol tweak.
\end{abstract}


\section{Introduction}
Quantum Key Distribution (QKD) protocols rely on both quantum and classical communication.
In particular, they require the use of authenticated classical channels to ensure the integrity of exchanged messages. A substantial body of work addresses this requirement, including information-theoretically secure authentication schemes~\cite{wegman_new_1981}, performance-optimized constructions ~\cite{krawczyk-LFSR-1994,kiktenko_lightweight_2020}, and formulations within composable security frameworks~\cite{portmann_key_2014}. Typically, in most existing QKD security proofs (see for instance \cite{tomamichel_largely_2017,mizutani2025protocolleveldescriptionselfcontainedsecurity}), the classical authentication is assumed to behave “honestly”: the authenticated channel never aborts, and all classical messages are delivered faithfully with their original timing preserved.
We refer to this as the honest authentication setting.
While this assumption simplifies security analyses, such an idealized channel cannot be realized in practice.

What can be constructed in practice is an authenticated channel that is close in functionality to one which either transmits each message faithfully or delivers a special symbol $\authabort$ to the receiving party, indicating that authentication has failed \cite{portmann_security_2022,portmann_key_2014}.
An authentication abort occurs, for example,  when the authentication tag attached to a message does not match the expected value, which can happen if the message has been modified or if an adversary attempts to inject a new message without the correct tag.
In either case, the receiver discards the message and registers $\authabort$.\footnote{In this work, we use the terms sender and receiver to denote parties sending and receiving that particular message. This has nothing to do with the sending and receiving of the quantum signals of the QKD protocol.}
Furthermore, an adversary (Eve) is permitted to delay, block, or reorder messages, or perform other timing-related manipulations, some of which may themselves result in an $\authabort$ being delivered to the receiver. We refer to this as the {\realistic}\footnote{By ``\realistic'' we mean a setting that is not itself perfectly achievable, but for which practical constructions can approximate the idealized functionality in the composable sense. See \cref{subsec:classicalcommunicationsmodel} for more discussion.} authentication setting.

This discrepancy leads to the following issues in the QKD analysis:
\begin{enumerate}
    \item First, the QKD protocol must now specify what happens when the authentication aborts. The natural choice here is to abort the protocol whenever authentication aborts (and attempt to communicate to the other party that one has aborted). 
    \item Second, only the receiving party is informed of the authentication aborts. Thus, Eve can 
    generally
    force \textit{one} party to abort in the QKD protocol while the other accepts (for instance by only interfering with the final message sent between the two parties). 
    \item Third, since the timing of messages may be modified, one can no longer assume a fixed ordering of actions performed by Alice and Bob. For instance, in the absence of synchronized clock assumptions in the security proof, Alice and Bob typically use messages to inform each other of completion of various operations in the protocol. If the timing of these messages is affected, then the ordering of actions performed by Alice and Bob is also affected.
\end{enumerate}  

These challenges arise in any setting where the authenticated classical channel may abort asymmetrically and where message timing is not rigidly preserved. Moreover, they require significant modifications to the security analysis:
\begin{itemize}
\item Due to Eve's ability to force asymmetric aborts in the {\realistic} authentication setting, the usual security definition of QKD as specified in \cite{portmann_security_2022,ferradini2025definingsecurityquantumkey,tupkary2025qkdsecurityproofsdecoystate,ben-or_universal_2004,renner_security_2005}, which only covers symmetric aborts, cannot be satisfied. This observation has been noted in prior works, see for instance, \cite[Section VII]{portmann_security_2022} 
\cite[Section 5.2.1]{tupkary2025qkdsecurityproofsdecoystate} \cite{ferradini2025definingsecurityquantumkey}. 
\item  Even if one chooses to disregard the problem of one-sided aborts, it remains important to recognize that existing QKD security analyses rely (often implicitly) on a fixed time ordering of classical communications, which is not guaranteed in practice. We include an illustrative example which highlights the importance of such considerations in Appendix~\ref{app:timing-attack}.
\end{itemize}

In this work, we address this gap as follows.
In \cref{subsec:securitydefinition}, we introduce the modified security definition for QKD protocols from Ref.~\cite{ferradini2025definingsecurityquantumkey}, which remains valid even when authentication can lead to receiver-side aborts.
This definition generalizes the usual trace-distance criterion by explicitly incorporating asymmetric abort events, while ensuring that if any party receives a key, it is secret from Eve.
In \cref{subsec:classicalcommunicationsmodel}, we specify a detailed model of interactive classical communication where authentication can result in one-sided aborts and where the adversary may modify the timing of classical messages (possibly resulting in authentication aborts).
We also briefly discuss how such a model can be implemented in practice. We then consider the scenario where we have an arbitrary ``core'' QKD protocol, which is followed by a short authentication post-processing (APP) step, described in \cref{subsec:APPprotocol}.
Our goal is to analyze the security of the combined core QKD + APP protocol in the {\realistic} setting where authentication can lead to asymmetric aborts and where message timing may be influenced by the adversary.
In \cref{subsec:reductionstatement}, we state our main result:  a reduction theorem showing that,
under mild and easily satisfied protocol conditions, if the core QKD protocol is secure under the honest authentication setting, then the combined protocol is secure under the {\realistic} authentication setting.This provides a clean separation between authentication and QKD security analysis, since one need not be concerned with authentication aborting or the timing of messages during classical communication while studying the security of the core QKD protocol. Moreover, it also retroactively lifts all prior QKD security proofs that were undertaken in the regime where the authentication was assumed to be honest to the more {\realistic} scenario, with the caveat that the protocol must now include the additional authentication post-processing step. The proof of this reduction is presented in \cref{sec:provingthereduction}.
Finally, in \cref{sec:delayedauthentication}, motivated by practical considerations of authentication key usage, we extend our analysis to the scenario of delayed authentication, where all classical communication during the core QKD protocol is undertaken using unauthenticated classical communication, and the entire communication transcript is authenticated at the end of the protocol. We  also discuss the trade-offs associated with this choice.
Concluding remarks are presented in \cref{sec:conclusion}. 

We note that some prior works have analyzed the security of QKD protocols in conjunction with a {\realistic} model of authentication, see, for example Ref.~\cite{kon_quantumauthenticated_2024}. However, the analysis in that work is tailored to a specific QKD protocol combined with a specific authentication protocol. In contrast, this work establishes a general result that applies to generic QKD protocols.

\subsection{Scope of modified security definition}

An important caveat is that in the modified security definition we study here, it is possible for the two parties in the QKD protocol to abort asymmetrically. In particular, this makes it a weaker security definition than the traditional one, meaning it may not be suitable for all scenarios in which users might previously have desired the traditional QKD security definition. Such scenarios would need to be carefully reconsidered on a case-by-case basis to determine whether the modified security definition still remains suitable, and it may well be the case that it does not suffice in some scenarios --- we do not aim to resolve such a wide-ranging task within the scope of this work, leaving it to be gradually considered over future work.

However, we wish to emphasize that our decision here to study a weaker security definition was not made arbitrarily. It might seem that a more natural approach to address the asymmetric abort issue would be to design a QKD protocol such that asymmetric aborts only happen with low probability, i.e.~with high probability both parties achieve consensus on whether to jointly accept or abort. Should it be possible to achieve this (with an obvious suggestion being to design the authentication protocol itself to avoid asymmetric aborts), this would of course immediately resolve that issue. Unfortunately, we are not aware of any such protocol in the literature, and there are indications that this should not be realizable --- in particular, such a goal appears extremely similar to resolving the Two Generals Problem, which is well-known to be a \emph{provably} impossible task, given only an untrusted public channel as a starting resource.\footnote{We stress that existing workarounds for the Two Generals Problem, such as sending additional message copies or using interactive multi-round protocols, do \emph{not} ensure consensus across a completely untrusted public channel --- they only do so under additional assumptions on the channel, such as limiting the fraction or probability of modified messages.}
While there are some subtle differences in our scenario compared to the Two Generals Problem, we believe that the essential ideas of the impossibility proof should carry over, and thus in this work we do not attempt this approach to resolve the asymmetric abort issue. Still, a more in-depth analysis may be of interest for future work. 

Therefore, the approach we propose in this work is meant to be a compromise position that acknowledges the current lack of prospects for avoiding asymmetric aborts (under standard assumptions), while still achieving what we believe to be a nontrivial cryptographic result. 
We note that while the final result is fairly intuitive, the task of establishing a rigorous proof of it is rather delicate, and requires a precise addressing of a number of aspects that we discuss in detail throughout the rest of this work.

Future work would be necessary in order to properly characterize the scope of viable applications for this result. As mentioned above, it may indeed turn out that allowing asymmetric aborts is unsuitable for some use cases --- however, we believe this should not be viewed as being ``due to'' the modified security definition, but rather as a critical flag highlighting that existing QKD protocols were \emph{already} invalid for such use cases, since they genuinely cannot avoid asymmetric aborts. For such use cases, it may be that the only solution would be to introduce additional ``non-standard'' assumptions to allow the parties to achieve consensus with high probability; as the choice of such assumptions would seem to very much depend on the use case, we leave that to be resolved in future work for each case as necessary.

We stress that this weaker definition raises no difficulty in terms of
\emph{composability} as an abstract concept. As claimed in Ref.~\cite[Sec.~VII.B]{portmann_security_2022}, the
definition still 
suffices to be a composably-secure construction of a particular choice of ideal resource, and this ideal resource still has the property that 
each party individually holds a key that has suitable secrecy properties against Eve's side
information. 
Such a resource can thus still be used in contexts where the main requirement is the secrecy of each party's key. The aspect that  requires deeper consideration: namely, whether the task at hand can tolerate the  possibility of an asymmetric abort, or whether it instead requires the two parties to reach agreement on whether a key was produced.

\section{Model and Security Definition} \label{sec:modeldefinition}
Let us begin by setting up some notation. 
We use uppercase letters such as $X, A, B$ to denote registers. We sometimes use bold uppercase letters registers such as $\bm{C},\bm{E}$ to  denote composite registers that may contain a large number of subregisters. We use $S_=(A)$ to denote the set of density operators on the register $A$, and $S_{\leq}(A)$ to denote the set of subnormalized density operators on the register $A$. Subscripts specify the registers on which the state exists; for example, $\rho_A$ denotes a state on register $A$. For a multipartite state $\rho_{ABC\cdots}$, we write $\rho_A$ to denote its marginal on register $A$. Conversely, for a state $\rho_A$, we use $\rho_{AB}$ to denote some extension of $\rho_A$ to an additional register $B$. We use $\CPTP(A,B)$ to denote the set of completely positive and trace preserving linear maps from registers $A$ to $B$. We use $\norm{\rho_A}_1$ to denote the Schatten one-norm of $\rho_A$. 

For a state $\rho \in \dop{\leq}(CQ)$ classical on $C$, written in the form
$\rho_{CQ} = \sum_c \ketbra{c}{c} \otimes \omega_c$ 
for some $\omega_c \in \dop{\leq}(Q)$,
and an event $\Omega$ defined on the register $C$, we will define a corresponding \emph{partial} state and \emph{conditional} state as, respectively,
\begin{align}
\rho_{\land\Omega} \defvar \sum_{c\in\Omega} \ketbra{c}{c} \otimes \omega_c, \qquad\qquad \rho_{|\Omega} \defvar \frac{\tr{\rho}}{\tr{\rho_{\land\Omega}}} \rho_{\land\Omega} = \frac{
\sum_{c} \tr{\omega_c}
}{\sum_{c\in\Omega} \tr{\omega_c}} \rho_{\land\Omega} .
\end{align}
We refer to $\rho_{\wedge \Omega}$ as the state $\rho$ being partial\footnote{One could instead also think of it as being ``subnormalized conditioned'' on the event $\Omega$.} on $\Omega$.

\subsection{Security Definition of QKD with asymmetric aborts} \label{subsec:securitydefinition}
We begin by specifying the security definition of QKD used in this work, adapted to account for asymmetric abort scenarios. We note that while this definition 
has not been explicitly analyzed within some composable security framework~\cite{maurer_abstract_2011,maurer_constructive_2012,maurer_indifferentiability_2016,portmann_security_2022,broadbent_2023,portmann_key_2014}, we expect that doing so should be straightforward, as claimed in Ref.~\cite{portmann_security_2022}. Let us consider the output state of a generic QKD protocol, defined on the registers $K_A K_B  \Esecdef$. Alice and Bob possess classical registers $K_A$ and $K_B$, which contain subregisters $K_A^{l_A}$ and $K_B^{l_B}$ that hold keys of length $l_A$ and $l_B$, respectively. We treat an abort by any party as that party storing a key of length $0$ in their register, represented by the special symbol $\bot$. The $\Esecdef$ register stores  all of Eve’s information at the end of the QKD protocol, and may include a copy of all classical communication that occurred during the protocol. The precise modeling of the authenticated classical communication, message timing, and one-sided authentication aborts is not essential for the security definition of QKD, which is only concerned with the output state of the QKD protocol, and will therefore be deferred to \cref{subsec:classicalcommunicationsmodel}. What matters for now is simply that the QKD protocol may output a state in which one party aborts while the other does not.

The output state of a generic QKD protocol can be written as \cite{ferradini2025definingsecurityquantumkey}:
\begin{equation} \label{cref:realoutputstate}
\rho^\text{real}_{K_A K_B  \Esecdef} \coloneq \bigoplus_{l_A, l_B \in \keyspace} \Pr(\OlAlB) \rho^\mathrm{real}_{K_A^{l_A} K_B^{l_B}  \Esecdef | \OlAlB},
\end{equation}
where $\OlAlB
$ denotes the event that Alice and Bob produce keys of lengths $l_A$ and $l_B$, respectively, and $\keyspace$ denotes the set of possible output key length combinations. The subregisters $K_A^{l_A}$ and $K_B^{l_B}$ store the keys of those specific lengths. As argued in Ref.~\cite{ferradini2025definingsecurityquantumkey}, 
we restrict the set of possible output key length combinations to the following: 
\begin{equation}
    \keyspace = \{  (l_A, l_B) \mid l_A = l_B \;\lor\; l_A = 0 \;\lor\; l_B = 0 \}.
\end{equation}

The ideal output state $\rho^\mathrm{ideal}_{K_A K_B  \Esecdef}$ is defined to be the one obtained by acting a map $\idealmap \in \CPTP( K_A K_B, K_A K_B)$ acting on the real output state $\rho^\mathrm{real}_{K_A K_B  \Esecdef}$ as
\begin{equation}
       \rho^\mathrm{ideal}_{K_A K_B  \Esecdef} \defvar \idealmap\left[\rho^\mathrm{real}_{K_A K_B  \Esecdef} \right],
\end{equation}
where $\idealmap$ is defined as performing  the following operations:
\begin{itemize}
    \item It looks at the length of the keys stored in registers $K_A, K_B$ to compute $l_A, l_B$. 
    \item It replaces the $K_A,K_B$ registers with the state 
\begin{equation}\label{eq:taukAkB}
\begin{aligned}
\tau^{l_A, l_B}_{K_A K_B} &\defvar
\begin{cases}
\displaystyle \frac{1}{2^{l_A}} \sum_{k \in \{0,1\}^{l_A}} \ketbra{kk}_{K_A K_B}, 
& \text{if } l_A = l_B, \\[1.2em]
\tau^{l_A}_{K_A} \otimes \tau^{l_B}_{K_B}
& \text{if } l_A \neq l_B,
\end{cases} \\[1.2em]
\tau^{l_A}_{K_A} &\coloneq \frac{1}{2^{l_A}} \sum_{k \in \{0,1\}^{l_A}} \ketbra{k}_{K_A}, \qquad 
\tau^{l_B}_{K_B} \coloneq \frac{1}{2^{l_B}} \sum_{k \in \{0,1\}^{l_B}} \ketbra{k}_{K_B} .
\end{aligned}
\end{equation}
    That is, if the output lengths are the same, the key registers are replaced with perfectly uniform identical keys of that length, independent of all other registers. If the output key lengths are not the same, the key registers are individually replaced with perfectly uniform keys of the corresponding lengths, independent of all other registers.
\end{itemize}

Thus intuitively, any key obtained from the ideal state is safe to use, regardless of symmetric or asymmetric aborts, since the key is always  independent of Eve's side-information registers. Note that we have
\begin{equation}
    \begin{aligned}
          \rho^\text{ideal}_{K_A K_B  \Esecdef}  &= \idealmap \left[ \rho^\text{real}_{K_A K_B  \Esecdef}   \right] \\
        &= \sum_{ \substack{l_A = l_B = l \\ l_A,l_B \in \keyspace}}  \Pr(\Omega_{l,l})  \tau^{l,l}_{K_A K_B} \otimes \rho^\text{real}_{  \Esecdef | \Omega_{l,l}} + \sum_{\substack{l_A \neq l_B \\ l_A,l_B \in \keyspace}} \Pr(\OlAlB) \tau^{l_A}_{K_A} \otimes \tau^{l_B}_{K_B} \otimes \rho^\text{real}_{  \Esecdef | \OlAlB}.
    \end{aligned}
\end{equation}
Moreover, $\idealmap$ acts independently on each combination of key length subregisters, i.e., it can be written as 
\begin{equation} \label{eq:idealmapdecomp}
    \idealmap = \bigoplus_{l_A, l_B} \idealmap^{(l_A,l_B)},  \qquad \text{ where $\idealmap^{(l_A,l_B)} \in \CPTP(K_A^{l_A} K_B^{l_B} , K_A^{l_A} K_B^{l_B} )$.}
\end{equation}

We now state the security definition, which will require us to talk about protocols and their corresponding output states. We therefore set up some notation first.
Let us consider a protocol \(\mathcal{P}\). What we will typically be concerned with is the set of output states produced by the protocol, which we denote by \(\mathcal{W}(\mathcal{P})\). This set of possible output states  depends on various assumptions ($\mathcal{W}$) under which we analyze the protocol. In this work, these assumptions will be related to the authenticated channel that is used during communication, and are explained in \cref{subsec:classicalcommunicationsmodel}.\footnote{For example, an assumption may be that Alice and Bob use unauthenticated classical communication, and Eve is allowed to tamper with classical messages. A different assumption may be that they use authenticated classical communication, and no tampering of the classical communication is allowed. These give rise to different sets of possible output states.}  Thus, these assumptions fix the set of possible attacks Eve can perform, and the protocol $\mathcal{P}$ along with Eve's attack together determine a channel mapping the relevant input states to output states. When we write \(\mathcal{W}(\mathcal{P})\), we refer to the set of all output states that can arise under all possible choices of Eve’s attack. Moreover, when we compose two protocols via \(\mathcal{P}_2 \circ \mathcal{P}_1\), the composed protocol is understood in the sense of channel composition, where the resulting channel is fixed by the protocol descriptions and Eve's attack on both protocols.  This level of formalism is sufficient for our purposes. 
 We can now state the QKD security definition. In this work, we will use $\coreQKDprotocol$, $\QKDprotocol$ to denote QKD protocols.

\begin{definition}[QKD Security with asymmetric aborts \cite{ferradini2025definingsecurityquantumkey}] \label{def:qkdsecurityasymmetric}
Let $\QKDprotocol$ be a QKD protocol, and let $\rho^\text{real}_{K_A K_B  \Esecdef}$ be the output state of the QKD protocol, and let $\mathcal{W}\left( \QKDprotocol \right)$ denote the set of possible output states of the QKD protocol.  Let $\rho^\text{ideal}_{K_A K_B  \Esecdef}$ be the ideal output state, obtained by acting the map $\idealmap$ on the actual output state. That is
\begin{equation} \label{eq:secdefrealideal}
    \begin{aligned}
        \rho^\text{real}_{K_A K_B  \Esecdef} &\coloneq \bigoplus_{l_A, l_B} \Pr(\OlAlB) \rho^\text{real}_{K_A^{l_A} K_B^{l_B}  \Esecdef| \OlAlB} \\
        \rho^\text{ideal}_{K_A K_B  \Esecdef}  &\defvar \idealmap \left[ \rho^\text{real}_{K_A K_B  \Esecdef}   \right]
    \end{aligned}
\end{equation}
Then, the QKD protocol is $\epssecure$-secure if, for all output states $\rho^\text{real}_{K_A K_B  \Esecdef} \in \mathcal{W}(\QKDprotocol)$, the following inequality is satisfied\footnote{Note that in Ref.~\cite{ferradini2025definingsecurityquantumkey}, the trace norm appearing in the security definition is not divided by $2$. In contrast, the typical security definition \cite{ben-or_universal_2004,portmann_security_2022} includes the explicit factor of $1/2$.  The definition used in Ref.~\cite{ferradini2025definingsecurityquantumkey} is deliberate and well motivated within that work; we stress that the difference amounts only to an overall factor of $2$ in the  security parameter.}:
\begin{equation}
    \tracedist{ \rho^\text{real}_{K_A K_B  \Esecdef} -  \rho^\text{ideal}_{K_A K_B  \Esecdef}} \leq \epssecure.
    \end{equation}
\end{definition}

\begin{remark} \label{remark:defgeneralize}
Note that this formulation generalizes the standard definition of QKD security \cite{ben-or_universal_2004,portmann_security_2022,ferradini2025definingsecurityquantumkey}. To see this, consider the setting where one is guaranteed to have $\Pr(\OlAlB) = 0$ whenever $l_A \neq l_B$ in the real output state (\cref{eq:secdefrealideal}). This corresponds to the typical scenario in QKD protocols with honest authentication, since the two honest parties can simply communicate their keylength decisions to ensure consistency. In this case, \cref{def:qkdsecurityasymmetric} reduces exactly to the standard definition of QKD security, since we have
\begin{equation}
    \begin{aligned}
        \rho^\mathrm{real}_{K_A K_B  \Esecdef} &=\bigoplus_{l} \Pr(\Omega_{l,l}) \rho^\mathrm{real}_{K_A^{l} K_B^{l}  \Esecdef| \Omega_{l,l}}, \\
        \rho^\mathrm{ideal}_{K_A K_B  \Esecdef}  &=\idealmap \left[ \rho^\mathrm{real}_{K_A K_B  \Esecdef}   \right] \\
        &=  \sum_{ l }  \Pr(\Omega_{l,l})  \tau^{l,l}_{K_A K_B} \otimes \rho^\mathrm{real}_{  \Esecdef | \Omega_{l,l}}. 
    \end{aligned}
\end{equation}
Note that this is merely an observation that often holds in QKD implementations under the honest authentication setting; none of the results of this work rely on this assumption.
\end{remark}
Throughout this work, we adopt the convention that the ideal and real states of various kinds are related analogously to \cref{eq:secdefrealideal}; that is, the ideal state is obtained by applying the map $\idealmap$ to the corresponding real state.
\subsection{Authenticated Classical Communication Model} \label{subsec:classicalcommunicationsmodel}
We will now explain the authenticated classical communications model we assume for the {\realistic} authentication setting. We stress that this model still contains idealized properties, in the sense that Eve has zero probability of faking messages without the authentication aborting. However, it is {\realistic} in the sense that it accounts for timing tampering and one-sided aborts. Moreover, we believe that an $\varepsilon_\mathrm{auth}$-close construction of such a channel is achievable within a composable framework (see \cref{subsubsec:implementingclassicalcomms} later for further discussion).

We begin by introducing notation to describe the sending and receiving of classical messages between Alice and Bob, which are mediated by Eve. For each message, we associate a register label from both the sender’s and receiver’s perspectives, and a global time (see \cref{fig:classicalcommmodel}). 
We emphasize that no assumptions are made about the alignment between the sender’s and receiver’s message ordering; these may differ in the presence of an active adversary.
Moreover, we do not assume that Alice and Bob share synchronized clocks, nor do they need to record the timing of messages during the protocol. The times introduced here are used only for the purpose of theoretical analysis and refer to the global time when these events occur. 
\begin{itemize}
    \item \textbf{Alice sending: } Alice sends messages  in registers $C^{(i)}_{A \rightarrow E}$, where the index $i$ denotes the ordering of messages from her perspective, i.e., in the order she sends them. We let $t^{(i)}_{A \rightarrow E}$ denote the time at which this message leaves Alice. 
     \item \textbf{Alice receiving: } Alice receives messages  $C^{(i)}_{E \rightarrow A}$, where the index $i$ denotes the ordering of messages from her perspective, i.e., in the order she receives them. We let $t^{(i)}_{E \rightarrow A}$ denote the time at which this message is received by Alice. 
    \item \textbf{Bob sending:} Bob sends messages in registers $C^{(i)}_{B \rightarrow E}$, where the index $i$ denotes the ordering of messages from Bob’s perspective, i.e., in the order he sends them. We let $t^{(i)}_{B \rightarrow E}$ denote the time at which this message leaves Bob. 

    \item \textbf{Bob receiving: } Bob receives messages  $C^{(i)}_{E \rightarrow B}$, where the index  $i$ denotes the ordering of messages from his perspective, i.e., in the order he receives them. We let $t^{(i)}_{E \rightarrow B}$ denote the time at which this message is received by Bob. 
\end{itemize}
Thus, each sent (received) message is indexed from the sender’s (receiver’s) point of view, and the time values reflect the true time at which these events occur (which are not known to Alice and Bob). We will modify this setting later in \cref{sec:delayedauthentication}.

We assume that the classical authenticated channel  between the two honest parties has the following properties:
\begin{enumerate}
    \item \textbf{Timing:} If the $i$th message is received \emph{before} the $i$th message was sent, then the received message is the special symbol $\authabort$. Formally,
    \begin{equation}
        \begin{aligned}
        &t^{(i)}_{A\rightarrow E} > t^{(i)}_{E \rightarrow B} \quad \implies \quad C^{(i)}_{E\rightarrow B} \text{ stores } \authabort, \quad \forall i.  \\
        &t^{(i)}_{B\rightarrow E} > t^{(i)}_{E \rightarrow A} \quad \implies \quad C^{(i)}_{E\rightarrow A} \text{ stores }\authabort, \quad \forall i.  
        \end{aligned}
    \end{equation}
We refer to the case where $t^{(i)}_{B\rightarrow E} \leq  t^{(i)}_{E \rightarrow A}$ and  $t^{(i)}_{A\rightarrow E} \leq t^{(i)}_{E \rightarrow B} $ holds  for all $i$ as ``relative time ordering being preserved''. For an explicit scenario where such considerations are critical, see Appendix~\ref{app:timing-attack}.

    \item \textbf{Modifying messages:} If $i$th message is received \textit{after} the $i$th message was sent, then the received message is either a copy of the sent message or it is an $\authabort$. Formally,
     \begin{equation}
        \begin{aligned}
        &t^{(i)}_{A\rightarrow E} \leq t^{(i)}_{E \rightarrow B} \quad \implies \quad \text{Either $C^{(i)}_{E\rightarrow B}$ and $C^{(i)}_{A\rightarrow E}$ store identical messages, or $C^{(i)}_{E\rightarrow B}$ stores  \authabort}, \quad \forall i.  \\
        &t^{(i)}_{B\rightarrow E} \leq t^{(i)}_{E \rightarrow A} \quad \implies \quad \text{Either $C^{(i)}_{E\rightarrow A}$ and $C^{(i)}_{B\rightarrow E}$ store identical messages, or $C^{(i)}_{E\rightarrow A}$ stores  \authabort}, \quad \forall i.  \\
        \end{aligned}
    \end{equation}

Our model implicitly imposes that for each message sent by some party, there exists \emph{some} finite time (possibly before the sending time) at which the other party receives some value, even if that value is simply \authabort: qualitatively, this means Eve is not allowed to delay messages ``indefinitely''. We discuss in \cref{remark:abortduetodelay} the potential measures that can be taken to realize this.

\end{enumerate}

\begin{remark}[Enforcing the finite delay duration] \label{remark:abortduetodelay}

Within this work, we do not attempt a full formalization of what would be required to construct the above model, as that requires addressing some technical points discussed further in \cref{subsubsec:implementingclassicalcomms}. However, it should not be difficult to at least impose the property that every sent message yields a ``received value'' at some finite time. Moreover, this can be enforced entirely locally by Alice and Bob, with no shared clock and no knowledge of send times. 

For example, each party could maintain a local timer for messages it is waiting to receive, and define a finite time window for each such message. The local timer can be started by a trigger that is common to both parties in the honest implementation, but which can be controlled by Eve in the adversarial case. If no correctly authenticated message arrives within the appropriate time window, the party records \authabort for that slot and closes it, discarding any later arrival.

In an honest implementation, the timing triggers can be synchronized using any standard mechanism, such as an initial handshake or conservative timeout estimates, so that unnecessary \authabort\ events occur only rarely. If Eve interferes with this synchronization --- for example, by delaying one party's timing trigger --- this may cause additional received values to become \authabort, but it does not invalidate the resource. Thus, synchronization serves only to improve the efficiency of honest executions and is not a security requirement. 
\end{remark}

For a given QKD protocol $\coreQKDprotocol$, we denote the set of output states possible in the above model for authenticated classical communication via $\worldreal(\coreQKDprotocol)$. In the honest authentication setting, we assume that Eve is not allowed to perform any operation that can result in an $\authabort$. That is, we are guaranteed to have:
\begin{equation}
    \begin{aligned}
        t^{(i)}_{A\rightarrow E} &\leq t^{(i)}_{E \rightarrow B} \qquad \qquad \forall i \\
        t^{(i)}_{B\rightarrow E} &\leq t^{(i)}_{E \rightarrow A} \qquad \qquad \forall i \\
        C^{(i)}_{E\rightarrow B} & \text{ and } C^{(i)}_{A\rightarrow E} \text{ store identical messages } \qquad \qquad \forall i \\  C^{(i)}_{E\rightarrow A} & \text{ and } C^{(i)}_{B\rightarrow E} \text{ store identical messages } \qquad \qquad \forall i
        \end{aligned}
\end{equation}
We denote the set of output states possible under this setting via $\worldhonest(\coreQKDprotocol)$.

\newcommand{\rowsep}{1.3}            
\tikzset{
  participant/.style={font=\small\bfseries},
  lifeline/.style={gray,dashed},
  message/.style={-Latex,thick},    
  attackbox/.style={draw,rounded corners,fill=blue!8,inner sep=4pt,font=\footnotesize}
}

\newcommand{\nextrow}[2]{\coordinate (#1) at ($(#2)+(0,-\rowsep)$);}
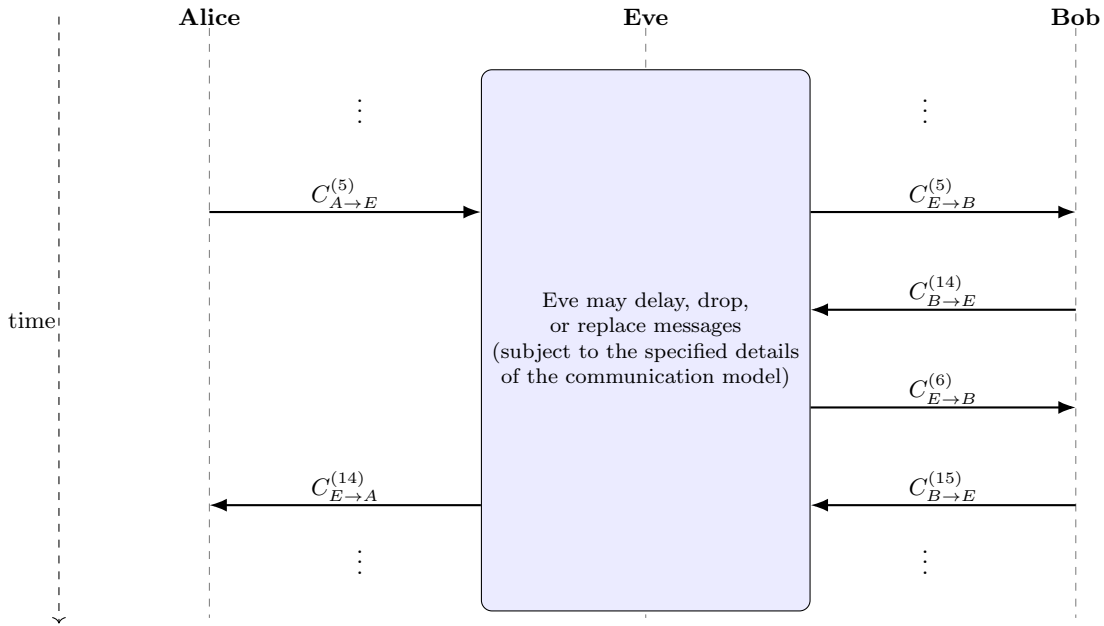
\begin{figure}[h!]
  \centering
  \begin{tikzpicture}[every node/.style={inner sep=1pt}]
    \node[participant] (Alice) {Alice};
    \node[participant] (Eve)   [right=5cm of Alice] {Eve};
    \node[participant] (Bob)   [right=5cm of Eve]   {Bob};

    \draw[lifeline] (Alice) -- coordinate (AliceEnd) ++(0,-8);
    \draw[lifeline] (Eve)   -- ++(0,-8);
    \draw[lifeline] (Bob)   -- coordinate (BobEnd) ++(0,-8);

    \coordinate (A0) at ($(Alice)+(0,-\rowsep)$);
    \coordinate (B0) at ($(Bob)+(0,-\rowsep)$);

    \node[attackbox,align=center,minimum height=7.2cm,anchor=north] (Ebox)
      at ($(Eve)+(0,-0.7)$) {Eve
      may delay, drop, \\
      or replace messages \\
      (subject to the specified details \\
      of the communication model)};

\node at ($ ($(Alice)!0.5!(A0)$) + ( 2,-0.5)$) {$\vdots$};
\node at ($ ($(Bob)!0.5!(B0)$)   + (-2,-0.5)$) {$\vdots$};

    \nextrow{A1}{A0}
    \nextrow{B1}{B0}

    \draw[message] (A1) -- node[midway,above] {$C_{A\to E}^{(5)}$} (Ebox.west |- A1);
    \draw[message] (Ebox.east |- A1) -- node[midway,above] {$C_{E\to B}^{(5)}$} (B1);

    \nextrow{A2}{A1}
    \nextrow{B2}{B1}

    \draw[message] (B2) -- node[midway,above] {$C_{B\to E}^{(14)}$} (Ebox.east |- B2);

    \nextrow{A3}{A2}
    \nextrow{B3}{B2}

    \draw[message] (Ebox.east |- A3) -- node[midway,above] {$C_{E\to B}^{(6)}$} (B3);

    \nextrow{A4}{A3}
    \nextrow{B4}{B3}

    \draw[message] (B4) -- node[midway,above] {$C_{B\to E}^{(15)}$} (Ebox.east |- B4);
    \draw[message] (Ebox.west |- B4) -- node[midway,above] {$C_{E\to A}^{(14)}$} (A4);

    \node at ($(A4)+(2,-0.5*\rowsep)$) {$\vdots$};
    \node at ($(B4)+(-2,-0.5*\rowsep)$) {$\vdots$};

\draw[->, dashed]
  ($(Alice)+(-2,0)$) -- ($(AliceEnd)+(-2,-4)$)
  node[midway, left] {time};

  \end{tikzpicture}
\caption{The {\realistic} authenticated  classical communication model used in this work. Messages pass through Eve, who may delay, drop, or substitute them with $\authabort$, subject to the constraints described in \cref{subsec:classicalcommunicationsmodel}. Time flows from top to bottom in the figure, which illustrates an example scenario: in earlier parts of the protocol (not shown in the figure), 4 messages have been sent from Alice to Bob, and 13 messages from Bob to Alice.
Eve does not interfere with Alice’s 5th message to Bob. However, she chooses to delay Bob’s 14th message. (Presumably, Alice does not send a new message during this period because she is waiting to receive one.) During the delay, Eve receives Bob’s 15th message and also delivers the 6th message to Bob. According to our communication model, this implies that $C^{(6)}_{E \rightarrow B}$ must be $\authabort$, since it was received before Alice sent her $6$th message.}
\label{fig:classicalcommmodel}
\end{figure}

\subsubsection{Implementing the authenticated communication model} \label{subsubsec:implementingclassicalcomms}
We note that there exist implementations of authenticated classical channels that closely approximate the functionality described by our model. One such implementation involves Alice and Bob sharing a pool of pre-distributed keys, which are used to authenticate each classical message using a message authentication scheme. As messages are sent and received, the parties iterate through their sending and receiving key pools. For example, they may employ Wegman-Carter authentication \cite{wegman_new_1981}, potentially with key recycling, as discussed in Ref.~\cite{portmann_key_2014}.

In such a setup, assuming the authentication keys remain secret, any attempt by Eve to modify a message will result in an invalid authentication tag with high probability, which causes the receiver to interpret the message as an $\authabort$. Moreover, if Eve attempts to deliver a message to the receiver \emph{before} the corresponding message has been sent, she will not have access to a valid message-tag pair. In this case as well, the receiver will reject the message as invalid (with high probability) and interpret it as $\authabort$. The same holds if, e.g., a pair of messages are swapped, since both messages would then result in $\authabort$s (with high probability). Note that in this description, there is no requirement for Alice and Bob to include the message index as a part of the message.

In practice, however, no implementation can perfectly realize the idealized model. There is always a small probability that Eve successfully forges a valid tag for a modified message.
To account for this, we could proceed in either of two ways. First, we could track this probability explicitly throughout our entire analysis, i.e.~noting at every step that there is some small probability of Eve forging the message without being detected, and writing the proof such that this event is explicitly tracked and accounted for. Alternatively, we could just perform our analysis entirely under the idealized model, and then rely on a separate proof that the implemented authentication protocol is a composably $\varepsilon_{\mathrm{auth}}$-secure construction (in some composable security framework) of the  authenticated channel described above --- given these, one can invoke composability to lift the security claims based entirely on the ideal case to the real implemented scenario. In this work, our analysis will be based on the latter approach, i.e.~we assume the parties have access to the authenticated channel with the functionality described earlier as a starting resource.

We acknowledge however that to our knowledge,  the existing literature on composable security currently does not contain an explicit construction of that resource \cite{portmann_key_2014,portmann_security_2022,broadbent_2023}; still, we believe such a construction to be achievable via similar arguments as in Ref.~\cite{portmann_key_2014}, and leave it as a point to be resolved in future work. In fact, if one disregards the technicalities introduced by message timing, the construction of a multi-use authenticated channel with one-sided aborts has already been demonstrated in Ref.~\cite{portmann_security_2022}.  

Finally, we point out that the classical communication model depends on the assumptions one makes on the adversary. In particular, it may be possible that such a model can be obtained by using computationally secure authentication schemes along with the assumption that Eve is unable to break such schemes during the time it takes to run a QKD protocol: however, we do not formally investigate this direction in this work.

\subsection{Authentication Post-Processing Protocol (APP)} \label{subsec:APPprotocol}
Recall our setting, where we have a core QKD protocol $\coreQKDprotocol$, followed by an Authentication Post-Processing protocol $\APPprotocol$. We refer to the resulting combined protocol as $\QKDprotocol = \APPprotocol \circ \coreQKDprotocol$. 
Our goal is to reduce the security of the combined protocol $\QKDprotocol$  (in the {\realistic} authentication setting) to the security of the core QKD protocol $\coreQKDprotocol$ alone (in the honest authentication setting). To achieve this, we design the Authentication Post-Processing protocol $\APPprotocol$ such that both parties abort whenever any message during the core QKD protocol $\coreQKDprotocol$ results in an $\authabort$, and such that it commutes with the map $\idealmap$. These properties are crucial in our proof of the reduction in \cref{sec:provingthereduction}. We will now specify the Authentication Post-Processing protocol.

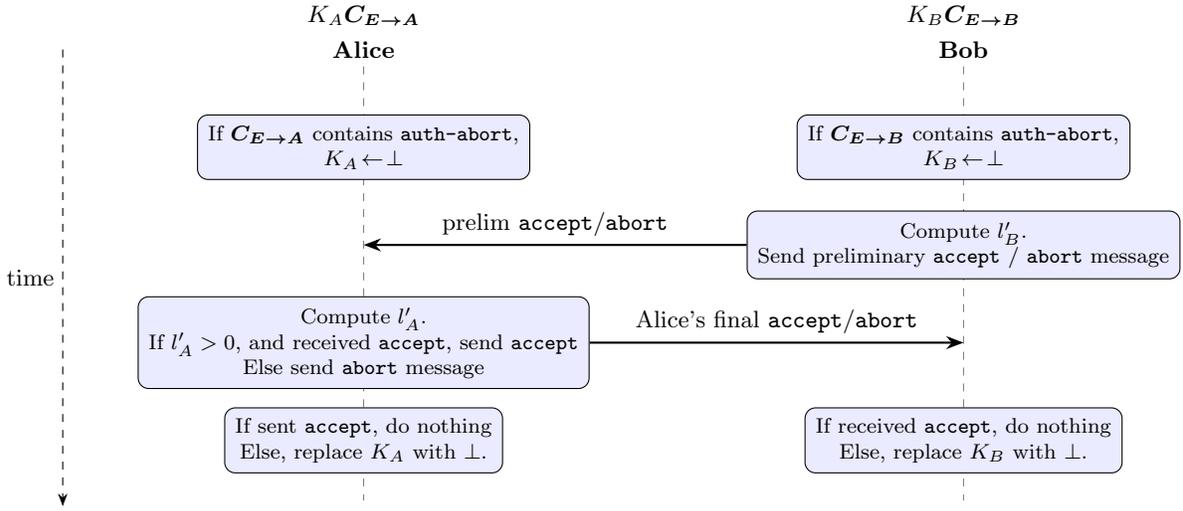
\begin{figure} 
\begin{tikzpicture}[
    >=Stealth,
    participant/.style={font=\small\bfseries},
    lifeline/.style  ={gray,dashed},
    message/.style   ={thick,-Stealth},
    localop/.style   ={draw,rectangle,rounded corners,
                       fill=blue!8,inner sep=4pt,
                       font=\footnotesize,align=center},
]

\def\rowsep{1.3} 

\node[participant] (Alice) {Alice};
\node[participant] (Bob)   [right=7cm of Alice] {Bob};

\node at ($(Alice)+(0,0.45)$) {$K_A \CAr $};
\node at ($(Bob)  +(0,0.45)$) {$K_B \CBr $};

\draw[lifeline] (Alice) -- ++(0,-6);
\draw[lifeline] (Bob)   -- ++(0,-6);

\coordinate (A0) at ($(Alice)+(0,-\rowsep)$);
\coordinate (B0) at ($(Bob)  +(0,-\rowsep)$);

\node[localop] (A1) at (A0) 
{If  $\CAr$ contains $\authabort$,\\
$K_A\!\leftarrow\!\bot$};
\node[localop] (B1) at (B0) 
{If  $\CBr$ contains $\authabort$,\\
$K_B\!\leftarrow\!\bot$};

\nextrow{B2}{B1}
\nextrow{A2}{A1}

\node[localop] (B2op) at (B2) {Compute $l_B'$. \\
Send preliminary $\accept$ / $\abort$ message};

\draw[message] (B2op) -- node[midway,above] {prelim \texttt{accept}/\texttt{abort}} (A2);


\nextrow{A3}{A2}
\nextrow{B3}{B2}

\node[localop] (A3op) at (A3) {Compute $l_A'$. \\
If $l_A' >0$, and received $\accept$, send $\accept$ \\
Else send $\abort$ message};

\draw[message] (A3op) -- node[midway,above] {Alice's final \texttt{accept}/\texttt{abort}} (B3);
\nextrow{A4}{A3}
\nextrow{B4}{B3}

\node[localop] (A4op) at (A4) {If sent $\accept$, do nothing \\
Else, replace $K_A$ with $\bot$.};

\node[localop] (B4op) at (B4) {If received $\accept$, do nothing \\
Else, replace $K_B$ with $\bot$.};

\draw[->, dashed]
  ($(Alice)+(-4,0)$) -- ($(AliceEnd)+(-4,-2)$)
  node[midway, left] {time};

\end{tikzpicture}
\caption{Schematic of \nameref{prot:authpp}  described in \cref{subsec:APPprotocol} without the pre-APP step. Alice and Bob first update their key registers based on whether they received an $\authabort$ in any of the prior communication. They then communicate their tentatively $\accept$ / $\abort$ decisions. They then perform a final update operation on their key registers depending on their final $\accept$ / $\abort$ decision. }
\label{fig:appprotocol}
\end{figure}

Let $\rho^\mathrm{real}_{K_A K_B \CfinalQKD \EfinalQKD}$ denote the final state obtained after the execution of the core QKD protocol $\coreQKDprotocol$, where $\CfinalQKD = \CAs \CBr \CBs \CAr$ collects all classical communication that occurred during the core protocol, and $\EfinalQKD$ denotes all of Eve’s side information, which may include a copy of the classical communication. During the authentication post-processing phase, Alice and Bob start with the state  $\rho^\mathrm{real}_{K_A K_B \CfinalQKD \EfinalQKD}$, perform some classical operations and communicate using registers $\CAsauth,\CBrauth,\CBsauth,\CArauth$ (see \cref{fig:appprotocol}). 
\vspace{2em}
\begin{prot}[AuthPP Protocol] \label{prot:authpp} 
Classical communication is undertaken in the registers
$\CAsauth, \CBrauth, \CBsauth, \CArauth$.
\begin{enumerate}[label=\textbf{APP~\arabic*}, ref=APP~\arabic*]

\item[pre-APP] \label{prot:preAPP} Alice sends a fixed, content-free message $\monofont{core-over}$ to Bob, indicating the termination of all of her steps for the core protocol. Bob waits for this dummy message to be received before proceeding to the steps of \nameref{prot:authpp}. We treat this as a special final step of the core protocol $\coreQKDprotocol$, with the message stored in $\CAs \CBr$. The reasoning behind this choice is explained in \cref{remark:specialstep} below. The state of the protocol is now 
$\rho^\mathrm{real}_{K_A K_B \CfinalQKD \EfinalQKD}$

\item \label{prot:authpp-alice} Alice checks whether any of her received messages in $\CAr$ is an \authabort. 
If she finds one, she replaces $K_A$ with $\bot$. 

\item \label{prot:authpp-bob} Bob checks whether any of his received messages in $\CBr$ is an \authabort. 
If he finds one, he replaces $K_B$ with $\bot$.

    \item \label{prot:authpp-bob2} Bob computes $l^\prime_B$ from $K_B$. If it is non-zero, he sends a preliminary \accept message to Alice. Otherwise, he sends an \abort message to Alice.
    
    \item \label{prot:authpp-alice2}Alice computes $l^\prime_A$ from $K_A$. If it is non-zero, \textit{and} she receives an \accept message from Bob, she sends her final \accept message to Bob. Else she sends \abort message to Bob.

       \item \label{prot:authpp-alice3}If Alice sent \accept message, she does nothing. If she sent \abort message, she replaces her key registers with $\bot$.
    
    \item \label{prot:authpp-bob3} If Bob receives an \accept message from Alice, he does nothing. If Bob receives either an \authabort\ or an \abort message, he replaces his key register $K_B$ with $\bot$. We let $l_B$ denote the final key length stored in $K_B$.
    
\end{enumerate}
\end{prot}
\vspace{2em}

%

Mathematically, the above protocols can be described as follows:
\begin{enumerate}
    \item In \cref{prot:authpp-alice} and \cref{prot:authpp-bob}, Alice and Bob simply replace their key registers with $\bot$s if they received an $\authabort$ during the prior protocol $\coreQKDprotocol$. This is described by  a map $\authmap \in \CPTP(\CfinalQKD K_A K_B,\CfinalQKD K_A K_B)$ (see \cref{remark:authprotsteps}).
    \item In \cref{prot:authpp-alice2,prot:authpp-bob2}, Alice and Bob first compute the key lengths $l^\prime_A, l^\prime_B$ stored in their key registers $K_A,K_B$ respectively at that point in time. Based on these values, they engage in two rounds of communications. For a given value of $l^\prime_A, l^\prime_B$, this can be described as a map (influenced by Eve) $\authcommmap^{(l^\prime_A,l^\prime_B)} \in \CPTP( \EfinalQKD, \Efinal \Cauth )$. The overall map is given by $\authcommmap \in \CPTP( K_A K_B \EfinalQKD, K_A K_B \Efinal \Cauth )$. Note that $\Cauth$ here contains separate messages that were sent and received, since Eve may attack this communication in any manner she desires, potentially modifying her quantum side-information as well.\footnote{Some messages have yet to be sent or received; however, for notational simplicity we let the map act on the entirety of $\Cauth$, with the understanding that it leaves registers meant to store future messages unchanged.}
    \item In \cref{prot:authpp-alice3,prot:authpp-bob3}, Alice and Bob  use the result of the communication in the previous step $\Cauth$ to determine their final $\accept$ / $\abort$ status. This is described as a map 
    $\authupdatemap \in \CPTP( K_A K_B \Cauth, K_A K_B \Cauth )$
\end{enumerate}
Let $\Onice$ be the event where neither Alice nor Bob receive an \authabort in $\CfinalQKD$ during $\coreQKDprotocol$ (including during the pre-APP step). In our communication model, this directly implies that all messages sent were received correctly by the receiving party, some time after they have been sent. In particular, it implies that Alice starts the pre-APP step and Bob starts \cref{prot:authpp-bob} only after all messages in the core protocol have been sent and received correctly, with no timing violations.  In this case, 
the final output state  is denoted by  \begin{equation} \label{eq:APPisaftercore}
\rho^\mathrm{real,final}_{K_A K_B \Cfinal \Efinal | \Onice} = \authupdatemap \circ \authcommmap \circ  \authmap  \left(\rho^\mathrm{real}_{K_A K_B \CfinalQKD \EfinalQKD | \Onice }\right),
\end{equation}
where $\Cfinal = \CfinalQKD \Cauth$, and we add a superscript `final' to the state to denote that this is the final output.

\begin{remark}[Role of the pre-APP step] \label{remark:specialstep}
The sole purpose of the pre-APP step above is to ensure that, conditioned on $\Onice$, Bob only sends the first message of  \nameref{prot:authpp} after the core protocol has terminated, in the sense that all messages from both parties have been sent and received. (The analogous requirement for Alice was already satisfied without this step). Note that local operations commute with Eve's operations, and hence we do not need to worry about when they were completed, and can focus our attention on the steps involving announcements. 

To see why we need this, consider 
a scenario where the last message of the core protocol is from Bob to Alice, and Eve chooses to delay it long enough that she can use Bob's message from \cref{prot:authpp-bob2} to influence her actions on this last message of the core protocol. Such a scenario is not accounted for in \cref{eq:APPisaftercore} which assumes that the \nameref{prot:authpp} steps occur after the core protocol has terminated, and presents a technical problem that must be addressed. This issue is remedied by the inclusion of the pre-APP step, which ensures that (conditioned on $\Onice$) the \nameref{prot:authpp} only makes announcements after the termination of the core QKD protocol.  Equivalently, the pre-APP step may be viewed as demanding that the final message of the core protocol be sent from Alice to Bob, since it can be included as a part of this last message.

We choose to regard the pre-APP step as part of the core protocol
$\coreQKDprotocol$ rather than as part of \nameref{prot:authpp}. This is purely a
matter of convenience for the analysis, in order to avoid having to separately account for the possibility of it
resulting in an \texttt{auth-abort} throughout the proof of our reduction. Finally, we emphasize that although the pre-APP step is equivalent to requiring
the final core message to be sent from Alice to Bob, this formulation lets us avoid 
imposing such a restriction on $\coreQKDprotocol$ directly in our subsequent statements. Moreover, since the pre-APP step involves a dummy message (which Eve knows already), its inclusion in the core protocol does not affect security statements made for that protocol in the honest authentication setting, which is the end point of our reduction. 
\end{remark}

\begin{remark} \label{remark:authprotsteps}
Note that the first two steps in the above protocol serve to ensure that if any classical communication resulted in an \authabort, then the receiving party stores a $\bot$  in the key register. In practice, one can simply enforce this property by appropriate design of the QKD protocol, i.e., by ensuring that the QKD protocol itself stores $\bot$ in the key register whenever \authabort is received. In such cases, we can omit \cref{prot:authpp-alice,prot:authpp-bob} from the authentication post-processing steps, as is done in Ref.~\cite{tupkary2026rigorouscompletesecurityproof}. In this work,  we wish to prove a statement that is agnostic to the nature of the QKD protocol implemented. Hence we include \cref{prot:authpp-alice,prot:authpp-bob} as explicit steps. 
\end{remark}

\subsection{Reduction Statement} \label{subsec:reductionstatement}
We are now ready to state the theorem that reduces the security analysis of QKD protocols to the setting in which authentication behaves honestly.

\begin{restatable}[Reduction of QKD security analysis to the honest authentication setting]{theorem}{reductionstatement}
\label{theorem:reductionstatement}

Let $\coreQKDprotocol$ be an arbitrary QKD protocol. Let $\APPprotocol$ be the \nameref{prot:authpp} described in \cref{subsec:APPprotocol}, executed after the core QKD protocol $\coreQKDprotocol$. Let $\QKDprotocol = \APPprotocol \circ \coreQKDprotocol$ denote the resulting QKD protocol. Let $\worldhonest(\coreQKDprotocol)$ denote the set of possible output states of  $\coreQKDprotocol$ in the honest authentication setting (see \cref{subsec:classicalcommunicationsmodel}). Let $\worldreal(\QKDprotocol)$ denote the set of possible output states of $\QKDprotocol$ in the {\realistic}  authentication setting (see \cref{subsec:classicalcommunicationsmodel}). 
Then, the $\epssecure$-security for all output states in $\worldhonest(\coreQKDprotocol)$ implies $\epssecure$-security for all output states\footnote{Note that any scenario in which Eve chooses to ignore or forget part of the public classical communication can be treated as one where she first records all communication and then traces out whatever she wishes at the end of the QKD protocol. Moreover, since QKD security analysis already allows Eve to retain all public communication, keeping the public communication register explicit and accessible to her is without loss of generality.} in $\worldreal(\QKDprotocol)$. That is,
\begin{equation}
    \begin{aligned}
         \tracedist{
\rho^\mathrm{real,hon}_{K_A K_B \CfinalQKD \EfinalQKD}
-
\rho^\mathrm{ideal,hon}_{K_A K_B \CfinalQKD \EfinalQKD}
} &\leq \epssecure \qquad \forall \rho^\mathrm{real,hon}_{K_A K_B \CfinalQKD \EfinalQKD} \in \worldhonest(\coreQKDprotocol) \\
          &\Downarrow \\
     \tracedist{ \rho^\mathrm{real,final}_{K_A K_B \Cfinal \Efinal} -  \rho^\mathrm{ideal,final}_{K_A K_B \Cfinal \Efinal}} &\leq \epssecure \qquad \forall \rho^\mathrm{real,final}_{K_A K_B \Cfinal \Efinal} \in \worldreal(\QKDprotocol).
    \end{aligned}
\end{equation}
\end{restatable}

\textit{Proof Idea.} The detailed proof of this theorem is presented in \cref{sec:provingthereduction}.
We define an event $\Onice$, corresponding to the case where no $\authabort$s are received during the execution of the core protocol $\coreQKDprotocol$. Under the communication model from \cref{subsec:classicalcommunicationsmodel}, any attempt by Eve to tamper with the message content or disturb the relative timing of messages leads to an $\authabort$. Thus, if we consider states partial on   $\Onice$, we may assume that the authentication behaves honestly. (Recall from the start of \cref{sec:modeldefinition}  that the state partial on $\Onice$ refers to the state that is conditioned on the event $\Onice$, but which is not re-normalized after the conditioning).

Since $\APPprotocol$ ensures that both parties abort whenever an $\authabort$ is received, it suffices to upper bound the trace distance from \cref{def:qkdsecurityasymmetric} for the output state of $\worldreal(\QKDprotocol)$ partial on $\Onice$. This is because if $\Onice$ does not occur, then both parties abort and the trace distance is zero by definition (see \cref{lemma:bothabort}).

Next, we observe that $\APPprotocol$ commutes with the ideal map $\idealmap$. We use this to show that it is sufficient to prove security for all states in $\worldreal(\coreQKDprotocol)$ partial on the event $\Onice$ (see \cref{lemma:commutationidealauth}).  The final step is showing that it then suffices to prove the security of the core QKD protocol $\coreQKDprotocol$ in the honest authentication setting. This is shown in \cref{lemma:reductionone,lemma:reductiontwo,lemma:reductionthree}.

\begin{remark}\label{remark:authnotetoreader}
\cref{theorem:reductionstatement} is quite general and serves as a bridge between the standard QKD security analyses performed under the assumption of honest authentication and the more {\realistic} setting where the authentication channel may be actively attacked, both in terms of message content and timing. Furthermore, we emphasize that it is entirely independent of the specific details of the QKD protocol being implemented, whether it is device-dependent or device-independent, prepare-and-measure or entanglement-based, etc. 
\end{remark}

\section{Proof of the Reduction \cref{theorem:reductionstatement}} \label{sec:provingthereduction}
In this section, we will provide a rigorous proof of \cref{theorem:reductionstatement}.

\subsection{Proving that \authabort in the core QKD protocol  results in both parties aborting}
We start by proving the following lemma, which states that if either Alice or Bob received an $\authabort$ in $\CfinalQKD$ during the core QKD protocol $\coreQKDprotocol$, then both parties will abort during authentication post-processing. We use $\Onice$ to denote the event that neither Alice nor Bob receive an \authabort in $\CfinalQKD$ during $\coreQKDprotocol$ (including the pre-APP step), and write $\Onice^\complement$ for its complement.

\begin{lemma} \label{lemma:bothabort}
Let $\Onice$ be the event where neither Alice nor Bob receive an \authabort in $\CfinalQKD$ during $\coreQKDprotocol$ (including the pre-APP step), and let $\rho^\mathrm{real,final}_{K_A K_B \Cfinal \Efinal}$ denote the final output state at the end of the full QKD protocol $\QKDprotocol$. Then the following equality holds
\begin{equation} \label{eq:niceeventequality}
 \idealmap \left[ \rho^\mathrm{real,final}_{K_A K_B \Cfinal \Efinal | \Onice^\complement} \right] \eqqcolon \rho^\mathrm{ideal,final}_{K_A K_B \Cfinal \Efinal | \Onice^\complement} = \rho^\mathrm{real,final}_{K_A K_B \Cfinal \Efinal | \Onice^\complement}.
\end{equation}
Therefore, 
\begin{equation} \label{eq:niceeventinequality}
\tracedist{\rho^\mathrm{real,final}_{K_A K_B \Cfinal \Efinal } - \rho^\mathrm{ideal,final}_{K_A K_B \Cfinal \Efinal }} =
    \tracedist{\rho^\mathrm{real,final}_{K_A K_B \Cfinal \Efinal \wedge \Onice} - \rho^\mathrm{ideal,final}_{K_A K_B \Cfinal \Efinal \wedge \Onice}} 
\end{equation}
\end{lemma}
\begin{proof}
    The proof of \cref{eq:niceeventequality} follows straightforwardly from the structure of the \nameref{prot:authpp} by considering the state  $\rho^\mathrm{real}_{K_A K_B \CfinalQKD \EfinalQKD | \Onice^\complement}$ and tracking it throughout the protocol.
    
    Let us suppose that Alice receives at least one $\authabort$. Then, \cref{prot:authpp-alice} replaces $K_A$ with $\bot$, and Alice announces \abort to Bob in \ref{prot:authpp-alice2}. Bob either  receives an $\abort$ or $\authabort$, and in either case, replaces his $K_B$ register with $\bot$ in \cref{prot:authpp-bob3}. The final output key length for both parties is $0$. 

Similarly, let us suppose that Bob receives at least one $\authabort$. Then, \cref{prot:authpp-bob} replaces $K_B$ with $\bot$, and Bob sends an $\abort$ message to Alice in \cref{prot:authpp-bob2}. This  leads to Alice sending an $\abort$ message to Bob in \cref{prot:authpp-alice2} and replacing her key register $K_A$ with $\bot$ in \cref{prot:authpp-alice3}. Thus, the final output key length for both parties is $0$.

The required \cref{eq:niceeventequality} follows by noting that $\idealmap$ acts as identity  when $l_A = l_B = 0$. Finally, \cref{eq:niceeventinequality} follows from \cref{eq:niceeventequality} via

\begin{equation} \begin{aligned}
\tracedist{\rho^\mathrm{real,final}_{K_A K_B \Cfinal \Efinal } - \rho^\mathrm{ideal,final}_{K_A K_B \Cfinal \Efinal }} 
&=   \tracedist{\rho^\mathrm{real,final}_{K_A K_B \Cfinal \Efinal \wedge \Onice} - \rho^\mathrm{ideal,final}_{K_A K_B \Cfinal \Efinal \wedge \Onice}}  \\
&+\tracedist{\rho^\mathrm{real,final}_{K_A K_B \Cfinal \Efinal \wedge \Onice^\complement} - \rho^\mathrm{ideal,final}_{K_A K_B \Cfinal \Efinal \wedge \Onice^\complement}}  \\
&=\tracedist{\rho^\mathrm{real,final}_{K_A K_B \Cfinal \Efinal \wedge \Onice} - \rho^\mathrm{ideal,final}_{K_A K_B \Cfinal \Efinal \wedge \Onice}}
    \end{aligned}
\end{equation}
where the first equality follows from the fact that the states conditioned on $\Onice$ and $\Onice^\complement$ live on orthogonal spaces, and the final equality follows from \cref{eq:niceeventequality}.
\end{proof}

\subsection{Reducing to security before authentication post-processing} \label{subsec:reducingtobeforeAPP}

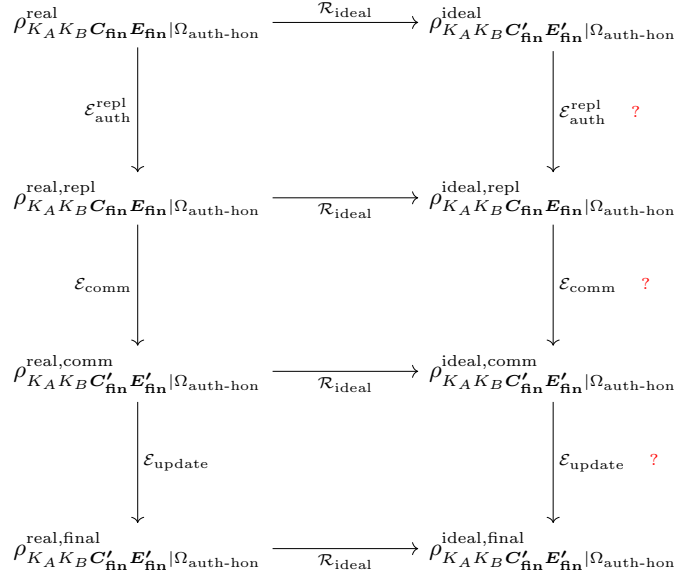
\begin{figure}[t!] 
\centering
\[
\begin{tikzcd}[row sep=5em, column sep=6em]
\rho^\text{real}_{K_A K_B \CfinalQKD \EfinalQKD | \Onice} \arrow[r, "\idealmap"] \arrow[d, "\authmap"'] 
      & \rho^\text{ideal}_{K_A K_B \CfinalQKD \EfinalQKD| \Onice} \arrow[d, "\authmap \quad {\color{red} ?}"] \\
\rho^\text{real,repl}_{K_A K_B \CfinalQKD \EfinalQKD| \Onice} \arrow[r, "\idealmap"'] \arrow[d, "\authcommmap"'] 
  &  \rho^\text{ideal,repl}_{K_A K_B \CfinalQKD \EfinalQKD| \Onice}  \arrow[d, "\authcommmap\quad {\color{red} ?}"] \\
\rho^\text{real,comm}_{K_A K_B \Cfinal \Efinal| \Onice}  \arrow[r, "\idealmap"'] \arrow[d,"\authupdatemap  "]
  & \rho^\text{ideal,comm}_{K_A K_B \Cfinal \Efinal| \Onice} \arrow[d,"\authupdatemap \quad {\color{red} ?}"] \\
  \rho^\text{real,final}_{K_A K_B \Cfinal \Efinal| \Onice}  \arrow[r, "\idealmap"'] 
  & \rho^\text{ideal,final}_{K_A K_B \Cfinal \Efinal| \Onice}
\end{tikzcd}
\]
\caption{A diagram illustrating transformations between real and ideal states evolving through the \nameref{prot:authpp}. The `{\color{red} $?$}' indicates the transformations that must be shown to be true in \cref{lemma:commutationidealauth}. The states go through the map $\authmap \in \CPTP(K_A K_B \CfinalQKD, K_A K_B \CfinalQKD)$ that replaces the key registers depending on $\CfinalQKD$. They then go through some communication steps given by $\authcommmap \in \CPTP(K_A K_B \EfinalQKD, K_A K_B \Efinal \Cauth) $ (influenced by Eve). Finally, Alice and Bob perform the final updates to their key registers described by $\authupdatemap \in \CPTP(K_A K_B \Cauth, K_A K_B \Cauth)$.}
\label{fig:comm-diagram}
\end{figure}

From \cref{eq:niceeventequality} we see that we only need to prove security for output states of $\QKDprotocol$ partial on $\Onice$. We would like to reduce the analysis to output states of $\coreQKDprotocol$ partial on $\Onice$.
We will show in \cref{lemma:commutationidealauth} (see also \cref{fig:comm-diagram}) that, conditioned on $\Onice$, the final real and ideal output states (at the end of $\QKDprotocol$) can be obtained by the action of $\authupdatemap \circ \authcommmap \circ \authmap $ on the real and ideal states at the end of the core QKD protocol $\coreQKDprotocol$.  Using the fact that the one-norm is non-increasing under CPTP maps, we can thus instead focus on the distance between the real and ideal output states \emph{before} the authentication post-processing. This distance will then be related to the usual security guarantee obtained under the assumption that authentication behaves honestly.

\begin{lemma}[Commutation of $\idealmap$ and \nameref{prot:authpp}] \label{lemma:commutationidealauth}
Let $\Onice$ denote the event where neither Alice nor Bob received an $\authabort$ in $\CfinalQKD$ during the core QKD protocol $\coreQKDprotocol$. Let $\rho^\mathrm{real}_{K_A K_B \CfinalQKD \EfinalQKD | \Onice}$ denote the real state at the end of the core QKD protocol  conditioned on $\Onice$. Let the following states denote its evolution through  \nameref{prot:authpp}, 
\begin{equation}
\begin{aligned}
   \rho^\mathrm{real,repl}_{K_A K_B \CfinalQKD \EfinalQKD| \Onice} &\coloneq \authmap \left[ \rho^\mathrm{real}_{K_A K_B \CfinalQKD \EfinalQKD| \Onice} \right], \\
    \rho^\mathrm{real,comm}_{K_A K_B \CfinalQKD \Cauth  \Efinal| \Onice} &\coloneq \authcommmap \left[ \rho^\mathrm{real,repl}_{K_A K_B \CfinalQKD \EfinalQKD| \Onice} \right], \\
     \rho^\mathrm{real,final}_{K_A K_B \Cfinal\Efinal| \Onice} &\coloneq \authupdatemap \left[ \rho^\mathrm{real,comm}_{K_A K_B \CfinalQKD \Cauth  \Efinal| \Onice} \right].
   \end{aligned}
\end{equation}
Define the corresponding ideal states by the action of the map $\idealmap$ on the real states, i.e 
\begin{equation}
\begin{aligned}
   \rho^\mathrm{ideal}_{K_A K_B \CfinalQKD \EfinalQKD| \Onice} &\coloneq \idealmap \left[   \rho^\mathrm{real}_{K_A K_B \CfinalQKD \EfinalQKD| \Onice} \right], \\
   \rho^\mathrm{ideal,repl}_{K_A K_B \CfinalQKD \EfinalQKD| \Onice} &\coloneq \idealmap \left[   \rho^\mathrm{real,repl}_{K_A K_B \CfinalQKD \EfinalQKD| \Onice} \right],\\
   \rho^\mathrm{ideal,comm}_{K_A K_B \CfinalQKD \Cauth \Efinal| \Onice} &\coloneq \idealmap \left[   \rho^\mathrm{real,comm}_{K_A K_B \CfinalQKD \Cauth\Efinal| \Onice} \right], \\
     \rho^\mathrm{ideal,final}_{K_A K_B \Cfinal\Efinal| \Onice} &\coloneq \idealmap \left[  \rho^\mathrm{real,final}_{K_A K_B \Cfinal\Efinal| \Onice}  \right].
   \end{aligned}
\end{equation}
Then, the ideal states defined above are the same as those obtained by evolving  $ \rho^\mathrm{ideal}_{K_A K_B \CfinalQKD \EfinalQKD| \Onice}  $ through  the \nameref{prot:authpp}, i.e.,
\begin{equation}
\begin{aligned}
      \rho^\mathrm{ideal,repl}_{K_A K_B \CfinalQKD \EfinalQKD| \Onice} &= \authmap \left[ \rho^\mathrm{ideal}_{K_A K_B \CfinalQKD \EfinalQKD| \Onice} \right], \\
    \rho^\mathrm{ideal,comm}_{K_A K_B \CfinalQKD \Cauth  \Efinal| \Onice} &= \authcommmap \left[ \rho^\mathrm{ideal,repl}_{K_A K_B \CfinalQKD \EfinalQKD| \Onice} \right], \\
     \rho^\mathrm{ideal,final}_{K_A K_B \Cfinal\Efinal| \Onice} &= \authupdatemap \left[\rho^\mathrm{ideal,comm}_{K_A K_B \CfinalQKD \Cauth  \Efinal| \Onice} \right].
     \end{aligned}
\end{equation} 
\end{lemma}

Note that this lemma also holds when conditioning on $\Onice^\complement$ instead of $\Onice$. However, the proof in that case is considerably more cumbersome, and we do not require it here. Recall that $\Cfinal = \CfinalQKD \Cauth$.
\begin{proof}
The proof consists of some straightforward though fairly cumbersome algebra. 
We wish to show that the three vertical arrows on the right hand side of \cref{fig:comm-diagram} are satisfied. We also note that none of the maps in $\APPprotocol$ ever affect the register $\CfinalQKD$ (they only read from the register), thus the event $\Onice$ can be defined at the start, and is ``preserved'' throughout the evolution of the real or ideal state. 

\paragraph*{First arrow:} For the first vertical arrow, we simply note that if $\Onice$ occurs, then the map $\authmap$ acts identically on the input state. Thus, 
\begin{equation} \label{eq:firstarrow}
  \authmap \circ \idealmap \left[  \rho^\text{real}_{K_A K_B \CfinalQKD \EfinalQKD | \Onice} \right] = \idealmap \circ \authmap  \left[  \rho^\text{real}_{K_A K_B \CfinalQKD \EfinalQKD | \Onice} \right]
\end{equation}
is trivially satisfied. 

\paragraph*{Second arrow:} For the second vertical arrow, we note that the $\authcommmap$ first looks at the key lengths stored in Alice's and Bob's registers, and then simply makes some announcements (and implements Eve's attacks). The $\idealmap$ map also simply replaces the key registers based on the key lengths. Neither map affects the key lengths stored in $K_A, K_B$. 
Let $\projkeymap \in \CPTP(K_A K_B, K_A K_B) $ denote the map that projects onto the subspace where the key registers store keys of lengths $l_A^\prime, l_B^\prime$. Then, 
since Alice's and Bob's announcements only depend on the length of their key registers $l_A^\prime$, $l_B^\prime$,  $\authcommmap$ has the following structure:
\begin{equation} \label{eq:authproofeq1}
    \authcommmap = \sum_{l^\prime_A , l^\prime_B}   \projkeymap \otimes \authcommmap^{(l_A^\prime, l_B^\prime)},
\end{equation}
 where $\authcommmap^{(l_A^\prime, l_B^\prime)} \in \CPTP(\CfinalQKD \EfinalQKD, \CfinalQKD \Cauth \Efinal)$ \footnote{It actually leaves the $\CfinalQKD$ register untouched.}, and does not act on the $K_A K_B$ registers. Moreover, the $\idealmap$ map also has a similar structure (\cref{eq:idealmapdecomp}), namely 
 \begin{equation} \label{eq:authproofeq2}
    \idealmap = \bigoplus_{l^\prime_A , l^\prime_B}  \idealmap^{(l_A^\prime, l_B^\prime)},  \qquad \text{ where $\idealmap^{(l^\prime_A,l^\prime_B)} \in \CPTP(K_A^{l^\prime_A} K_B^{l^\prime_B} , K_A^{l^\prime_A} K_B^{l^\prime_B} )$.}
\end{equation}
Then, we note that
\begin{equation} 
\begin{aligned}
\label{eq:authproofeq3}
   \authcommmap \circ \idealmap &= \sum_{l_A^\prime, l_B^\prime}  \projkeymap \circ \idealmap^{(l^\prime_A,l^\prime_B)}  \otimes \authcommmap^{(l_A^\prime, l_B^\prime)} \\
   &= \sum_{l_A^\prime, l_B^\prime}  \idealmap^{(l^\prime_A,l^\prime_B)} \circ  \projkeymap   \otimes \authcommmap^{(l_A^\prime, l_B^\prime)}  \\
   &= \idealmap \circ \authcommmap,
    \end{aligned}
\end{equation}
where all the equalities follow from the definition of the maps.  The required claim then follows from \cref{eq:authproofeq3}.

\paragraph*{Third arrow:} For the third vertical arrow, we use the fact that $\authupdatemap$ simply reads from the $\Cauth$ register and performs operations on the $K_A,K_B$ registers. Its operations depend on the values observed in the $\Cauth$ register. To proceed we will require some additional notation.

Let us use $\finalacceptabortevent{i}{j}$, where $i,j$ can either be $\top$ or $\bot$ to denote the final accept or abort decisions (and communication) undertaken by Alice and Bob as follows: 
\begin{itemize}
    \item If $i=\top$, Alice  does nothing in \cref{prot:authpp-alice3}. This corresponds to Alice sending a final \accept message in \cref{prot:authpp-alice2}.
    \item If $i=\bot$, Alice   replaces the key register with the $\bot$ value in \cref{prot:authpp-alice3}. This corresponds to Alice sending a final \abort message in \cref{prot:authpp-alice2}.
    \item If $j = \top$, Bob does nothing in \cref{prot:authpp-bob3}. This corresponds to him receiving an \accept message from Alice in \cref{prot:authpp-alice2}.
    \item If $j = \bot$, Bob replaces his key register with $\bot$ in \cref{prot:authpp-bob3}. This corresponds to him receiving either an \abort or \authabort message from Alice in \cref{prot:authpp-alice2}. 
\end{itemize}   
The $\idealmap$ map also performs operations on the $K_A K_B$ register, and its operations only depend on the lengths of the strings stored in $K_A,K_B$. Recall from \cref{subsec:APPprotocol} that we use $l_A^\prime, l_B^\prime$ to denote these lengths before the update operation, and $l_A, l_B$ to denote these lengths after the update operation.

Let us first consider the scenario where we first apply $\authupdatemap$ and then $\idealmap$. Thus, we will focus first on the state $ \rho^\mathrm{real,comm}_{K_A K_B \Cfinal \Efinal | \Onice \wedge \OlAlBprime}$. This state is a mixture of the following states which correspond to various final accept or abort decisions:
\begin{equation} \label{eq:commproofmixture}
\begin{aligned}
    \Big\{ &\rho^\mathrm{real,comm}_{K_A K_B \Cfinal \Efinal | \Onice \wedge \OlAlBprime\wedge \finalacceptabortevent{\top}{\top} }, \\
    &\rho^\mathrm{real,comm}_{K_A K_B \Cfinal \Efinal | \Onice \wedge \OlAlBprime\wedge\finalacceptabortevent{\bot}{\bot} }, \\
    &\rho^\mathrm{real,comm}_{K_A K_B \Cfinal \Efinal | \Onice \wedge \OlAlBprime\wedge\finalacceptabortevent{\top}{\bot} }, \\
    &\rho^\mathrm{real,comm}_{K_A K_B \Cfinal \Efinal | \Onice \wedge \OlAlBprime\wedge\finalacceptabortevent{\bot}{\top} } \Big\}.
    \end{aligned}
\end{equation}
(Note that for the \nameref{prot:authpp} considered in this work, we can never have the $\finalacceptabortevent{\bot}{\top}$ occur. However, we leave it in our analysis to enable this proof to be easily adapted to variations of \nameref{prot:authpp}).
From the actions of the update map (see \cref{prot:authpp-alice3,prot:authpp-bob3}), it is straightforward to verify that  
\begin{equation}
    \authupdatemap \circ \idealmap \left[ \cdot \right] = \idealmap \circ \authupdatemap \left[ \cdot \right] \qquad \text{where $\cdot$ is any state from \cref{eq:commproofmixture}}.
\end{equation}
That is, the required commutation holds conditioned on any combination of final \accept / \abort outcomes. This implies 

\begin{equation}
    \authupdatemap \circ \idealmap \left[ \rho^\mathrm{real,comm}_{K_A K_B \Cfinal \Efinal | \Onice \wedge \OlAlBprime}\right] = \idealmap \circ \authupdatemap \left[ \rho^\mathrm{real,comm}_{K_A K_B \Cfinal \Efinal | \Onice \wedge \OlAlBprime} \right] \qquad \forall l_A^\prime ,l_B^\prime.
\end{equation}
Since the above equation holds for all values of the intermediate key lengths $l_A^\prime, l_B^\prime$, we have the required claim: 

\begin{equation} \label{eq:arrow3}
    \authupdatemap \circ \idealmap \left[ \rho^\mathrm{real,comm}_{K_A K_B \Cfinal \Efinal | \Onice }\right] = \idealmap \circ \authupdatemap \left[ \rho^\mathrm{real,comm}_{K_A K_B \Cfinal \Efinal | \Onice } \right].
\end{equation}

Thus, the stated Lemma follows from \cref{eq:firstarrow,eq:arrow3,eq:authproofeq3}.
This concludes the proof.
\end{proof}

This commuting property allows us to focus on the distance between the real and ideal states \emph{before} \nameref{prot:authpp}, rather than after. This can be formalized using the following corollary.

\begin{corollary} \label{corr:reduction}
Let $\coreQKDprotocol$, $\APPprotocol$, and $\QKDprotocol$ be protocols such that the overall QKD protocol is given by $\QKDprotocol = \APPprotocol \circ \coreQKDprotocol$, where $\APPprotocol$ denotes a post-processing routine executed after the core QKD protocol $\coreQKDprotocol$. Assume that classical communication in the {\realistic} and honest setting behaves as specified in \cref{subsec:classicalcommunicationsmodel}, and suppose that $\APPprotocol$ is as described in \nameref{prot:authpp}. That is, consider the same setup as in \cref{theorem:reductionstatement}. Then, the $\epssecure$-security for all output states in $\worldreal(\coreQKDprotocol)$ partial on the event $\Onice$, implies $\epssecure$-security for all output states in $\worldreal(\QKDprotocol)$. That is 
  \begin{equation}
      \begin{aligned}
    \tracedist{\rho^\mathrm{real}_{K_A K_B \CfinalQKD \EfinalQKD \wedge \Onice }  - \rho^\mathrm{ideal}_{K_A K_B \CfinalQKD \EfinalQKD \wedge \Onice} }  &\leq \epssecure \qquad \forall \rho^\mathrm{real}_{K_A K_B \CfinalQKD \EfinalQKD } \in  \worldreal(\coreQKDprotocol) \\
    &\Downarrow \\
     \tracedist{\rho^\mathrm{real,final}_{K_A K_B \Cfinal \Efinal }  - \rho^\mathrm{ideal,final}_{K_A K_B \Cfinal \Efinal } }  &\leq \epssecure \qquad \forall \rho^\mathrm{real,final}_{K_A K_B \Cfinal \Efinal } \in  \worldreal(\QKDprotocol) 
      \end{aligned}
  \end{equation}
\end{corollary}
\begin{proof}
   For any output state $\rho^\mathrm{real,final}_{K_A K_B \Cfinal \Efinal } \in \worldreal(\QKDprotocol)$ of the full QKD protocol, consider the corresponding state $\rho^\mathrm{real}_{K_A K_B \CfinalQKD \EfinalQKD } \in \worldreal(\coreQKDprotocol)$ obtained at the end of the core QKD protocol $\coreQKDprotocol$. Then, we have
    \begin{equation}
        \begin{aligned}
         \tracedist{\rho^\mathrm{real,final}_{K_A K_B \Cfinal \Efinal } - \rho^\mathrm{ideal,final}_{K_A K_B \Cfinal \Efinal }} &= 
    \tracedist{\rho^\mathrm{real,final}_{K_A K_B \Cfinal \Efinal \wedge \Onice} - \rho^\mathrm{ideal,final}_{K_A K_B \Cfinal \Efinal \wedge \Onice}} \\
    &=  \tracedist{ \authupdatemap \circ \authcommmap \circ \authmap \left[ \rho^\mathrm{real}_{K_A K_B \CfinalQKD \EfinalQKD \wedge \Onice} - \rho^\mathrm{ideal}_{K_A K_B \CfinalQKD \EfinalQKD\wedge \Onice} \right]} \\
    &\leq \tracedist{ \rho^\mathrm{real}_{K_A K_B \CfinalQKD \EfinalQKD \wedge \Onice} - \rho^\mathrm{ideal}_{K_A K_B \CfinalQKD \EfinalQKD\wedge \Onice} } 
\end{aligned}\end{equation}
where we use \cref{lemma:bothabort} for the first line, \cref{lemma:commutationidealauth} for the second line, and the fact that the one-norm cannot increase under the action of CPTP maps for the final line. This suffices to prove the required claim.  \end{proof}

Thus, \cref{corr:reduction} allows us to restrict our attention to bounding 
\begin{equation} \label{eq:targetofreduction}
  \tracedist{\rho^\mathrm{real}_{K_A K_B \CfinalQKD \EfinalQKD \wedge \Onice }  - \rho^\mathrm{ideal}_{K_A K_B \CfinalQKD \EfinalQKD \wedge \Onice} }  \leq \epssecure \qquad \forall \rho^\mathrm{real}_{K_A K_B \CfinalQKD \EfinalQKD } \in  \worldreal(\coreQKDprotocol)
\end{equation}
which is the distance between the real and ideal states after the core QKD protocol (partial on the event $\Onice$).

\subsection{Reducing to scenario where authentication satisfies honest behaviour}
The task now is to show that if the QKD protocol is analyzed under the assumption that authentication follows honest behaviour, then this analysis applies to output states (partial on $\Onice$) in the setting where authentication is not assumed to be honest. Under the communication model defined in \cref{subsec:classicalcommunicationsmodel}, the event $\Onice$ coincides with the event that authentication behaved honestly. Indeed, any attempt to tamper with the contents of a message or violate the temporal ordering of messages triggers an $\authabort$ (i.e., $\Onice^\complement$). Thus, it is natural to relate the two scenarios. 

While this may initially seem straightforward, making the connection formal is non-trivial. One might intuitively expect that conditioning on $\Onice$ should allow us to restrict attention to the honest-authentication setting. However, this naive reasoning fails upon realizing that the set of operations available to Eve in the real setting is strictly \emph{larger} than that in the honest authentication setting.\footnote{For example, consider an operation in which Eve --- perhaps probabilistically, or via a unitary interaction between her quantum side information and the classical signals --- preemptively sends messages to one party in order to induce that party to advance to the next stage of the protocol earlier than intended (see Appendix~\ref{app:timing-attack} for an example of such a scenario). Such an operation is simply not \emph{allowed} in the honest authentication setting, therefore the resulting output state cannot be identified in $\worldhonest(\coreQKDprotocol)$.
} Consequently, the set of possible output states in the {\realistic} authentication setting $\worldreal(\coreQKDprotocol)$ is strictly \emph{larger} than the set of possible output states in the honest authentication setting $\worldhonest(\coreQKDprotocol)$ (even if we consider partial states on $\Onice$).
To circumvent this problem, we introduce a \emph{virtual authentication setting}  (i.e., a new set of assumptions describing a virtual authentication scenario) which acts as a bridge that allows us to relate the two settings. This technical step is essential for our reduction argument that follows. Thus, the series of reductions that we build is (informally) described by:
 \begin{equation} \label{eq:reductionpath}
\begin{aligned}
   & \worldreal(\QKDprotocol) \xrightarrow{\cref{lemma:bothabort}}\worldreal(\QKDprotocol)_{\wedge \Onice} \xrightarrow{\cref{corr:reduction}} \worldreal(\coreQKDprotocol)_{\wedge \Onice}  \xrightarrow{\cref{lemma:reductionone}} \\
   &\worldvirtual(\coreQKDprotocol)_{\wedge \Onice} 
    \xrightarrow{\cref{lemma:reductiontwo}} 
    \worldvirtual(\coreQKDprotocol) \xrightarrow{\cref{lemma:reductionthree}}
    \worldhonest(\coreQKDprotocol),
    \end{aligned}
\end{equation}
where the set of output states in the virtual setting is denoted using $\worldvirtual$ and described shortly.

\subsubsection{Describing the virtual authentication setting}  \label{subsec:virtualsetting}
 In the virtual authentication setting, Eve still has the ability to implement any attack she wants on the classical channel; in particular, she controls the message timings $t^{(i)}_{E \rightarrow A}, t^{(i)}_{E \rightarrow B}$ and the contents of $C^{(i)}_{E \rightarrow A},C^{(i)}_{E \rightarrow B}$. Thus, the set of possible operations she can do is \emph{exactly the same} as in the {\realistic} authentication setting. However, the virtual setting is designed such that it is equivalent to the {\realistic} authentication setting whenever event $\Onice$ occurs, through the use of special registers $\bm{\Ccorr}$. The virtual authentication setting is described as follows:
\begin{enumerate}
\item Eve's attack results in the registers $\CAs,\CAr,\CBs,\CBr$ being sent and received in the exact same way as the {\realistic} authentication setting (see  \cref{subsec:classicalcommunicationsmodel}). As before,  each party labels their outgoing (incoming)
messages according to the order in which the messages are sent (received).
    \item \textbf{Timing:} If the $i$th message is received \textit{before} the $i$th message was sent, the receiving party gets the correct message in additional registers $\bm{\Ccorr}$, at some time after the message was sent. Stated formally:
 \begin{equation}
        \begin{aligned}
        &t^{(i)}_{A\rightarrow E} > t^{(i)}_{E \rightarrow B} \quad \implies \text{Bob receives correct message in $\Ccorr^{(i)}_{E \rightarrow B}$ at time after  $t^{(i)}_{A\rightarrow E}$} , \quad \forall i. \\
       &t^{(i)}_{B\rightarrow E} > t^{(i)}_{E \rightarrow A} \quad \implies \text{Alice receives correct message in $\Ccorr^{(i)}_{E \rightarrow A}$ at time after $t^{(i)}_{B\rightarrow E}$}, \quad \forall i.
        \end{aligned}
    \end{equation}

   \item \textbf{Modifying messages:} If $i$th message is received \textit{after} the $i$th message was sent, then the  receiving party gets the correct message in additional registers $\bm{\Ccorr}$, at the same time as the actual message is received. Stated formally:
     \begin{equation}
        \begin{aligned}
        &t^{(i)}_{A\rightarrow E} \leq t^{(i)}_{E \rightarrow B} \quad \implies \quad \text{Bob receives the correct message in $\Ccorr^{(i)}_{E \rightarrow B}$ at time $t^{(i)}_{E \rightarrow B}$}, \quad \forall i.  \\
        &t^{(i)}_{B\rightarrow E} \leq t^{(i)}_{E \rightarrow A} \quad \implies \quad \text{Alice receives the correct message in $\Ccorr^{(i)}_{E \rightarrow A}$ at time $t^{(i)}_{E \rightarrow A}$}, \quad \forall i.
        \end{aligned}
    \end{equation}
\item Alice and Bob use the registers $\bm{\Ccorr_{E \rightarrow A}},\bm{\Ccorr_{E \rightarrow B}}$ for the received messages, and implement decisions based on the correct copies they receive. They entirely ignore the messages received in $\CAr, \CBr$.
\end{enumerate}
We use $\worldvirtual(\coreQKDprotocol)$ to denote the set of output states that can be obtained in the virtual setting for the given protocol $\coreQKDprotocol$. We will now prove a series of lemmas that reduce the $\epssecure$-security statement we wish to prove (\cref{eq:targetofreduction}) to $\epssecure$-security of all states in $\worldhonest(\coreQKDprotocol)$.  We remark that, strictly speaking, the protocols in the virtual setting differ slightly from those in the {\realistic} authentication setting, as they make use of distinct classical communication registers (the virtual setting uses the corresponding special registers). However, since these registers are equivalent by construction, and Alice and Bob  perform the same operations based on the received communication in both the {\realistic} and virtual authentication settings, we do not label these protocols differently across the two.

\subsubsection{Reducing security statements}
We first reduce the security analysis to states in $\worldvirtual(\coreQKDprotocol)$ partial on $\Onice$.
\begin{lemma} \label{lemma:reductionone}
        Consider the same setup as in \cref{theorem:reductionstatement,corr:reduction}. Then, the $\epssecure$-security for all output states in $\worldvirtual(\coreQKDprotocol)$ partial on $\Onice$ implies $\epssecure$-security for all output states in $\worldreal(\coreQKDprotocol)$ partial on $\Onice$. That is 
  \begin{equation}
      \begin{aligned}
    \tracedist{\rho^\mathrm{real,virt}_{K_A K_B \CfinalQKD \bm{\Ccorr} \EfinalQKD \wedge \Onice }  - \rho^\mathrm{ideal,virt}_{K_A K_B \CfinalQKD \bm{\Ccorr}  \EfinalQKD \wedge \Onice} }  &\leq \epssecure \qquad \forall \rho^\mathrm{real,virt}_{K_A K_B \CfinalQKD \bm{\Ccorr} \EfinalQKD } \in  \worldvirtual(\coreQKDprotocol) \\
    &\Downarrow \\
     \tracedist{\rho^\mathrm{real}_{K_A K_B \CfinalQKD \EfinalQKD \wedge \Onice}  - \rho^\mathrm{ideal}_{K_A K_B \CfinalQKD \EfinalQKD \wedge \Onice} }  &\leq \epssecure \qquad \forall \rho^\mathrm{real}_{K_A K_B \CfinalQKD \EfinalQKD  } \in  \worldreal(\coreQKDprotocol) 
      \end{aligned}
  \end{equation}
\end{lemma}
\begin{proof}
    Fix any attack by Eve, and consider the corresponding state $\rho^\mathrm{real}_{K_A K_B \CfinalQKD \EfinalQKD} \in \worldreal(\coreQKDprotocol)$:
    \begin{equation}
        \rho^\mathrm{real}_{K_A K_B \CfinalQKD \EfinalQKD} = \rho^\mathrm{real}_{K_A K_B \CfinalQKD \EfinalQKD \wedge \Onice }+ \rho^\mathrm{real}_{K_A K_B \CfinalQKD \EfinalQKD \wedge \Onice^\complement}.
    \end{equation}
    For the same attack, consider the corresponding state $\rho^\mathrm{real,virt}_{K_A K_B \CfinalQKD  \bm{\Ccorr} \EfinalQKD} \in \worldvirtual(\coreQKDprotocol)$ in the virtual setting:
    \begin{equation}
        \rho^\mathrm{real,virt}_{K_A K_B \CfinalQKD  \bm{\Ccorr}\EfinalQKD} = \rho^\mathrm{real,virt}_{K_A K_B \CfinalQKD  \bm{\Ccorr} \EfinalQKD \wedge \Onice }+ \rho^\mathrm{real,virt}_{K_A K_B \CfinalQKD  \bm{\Ccorr} \EfinalQKD \wedge \Onice^\complement}.
        \end{equation}
Recall that the virtual setting is the one where the correct messages are received by both parties in the registers $\bm{\Ccorr}$, and Alice and Bob use these values in the protocol. In the event $\Onice$, the actual messages (received in $\CfinalQKD$) are exactly the same as the  correct messages received in $\bm{\Ccorr}$, and are received at exactly the same time as the correct messages in $\bm{\Ccorr}$. Thus, these two registers are  equivalent, and lead to Alice and Bob performing the same operations at all times (conditioned on the event $\Onice$). Thus, we have 
\begin{equation}
\begin{aligned}
    \Tr_{\bm{\Ccorr}} \left[\rho^\mathrm{real,virt}_{K_A K_B \CfinalQKD  \bm{\Ccorr} \EfinalQKD \wedge \Onice}  \right] &= \rho^\mathrm{real}_{K_A K_B \CfinalQKD   \EfinalQKD \wedge \Onice} \\
     \Tr_{\bm{\Ccorr}} \left[\rho^\mathrm{ideal,virt}_{K_A K_B \CfinalQKD  \bm{\Ccorr} \EfinalQKD \wedge \Onice}  \right] &= \rho^\mathrm{ideal}_{K_A K_B \CfinalQKD   \EfinalQKD \wedge \Onice}
    \end{aligned}
\end{equation}
where the second equality follows from the first equality, and the fact that $\idealmap$ does not act on $\bm{\Ccorr}$.
With these two equalities, we obtain the desired claim by using the fact that the one-norm decreases under the action of CPTP maps, as follows:
\begin{equation}
\begin{aligned}
&\tracedist{\rho^\mathrm{real}_{K_A K_B \CfinalQKD \EfinalQKD \wedge \Onice}  - \rho^\mathrm{ideal}_{K_A K_B \CfinalQKD \EfinalQKD \wedge \Onice} }  \\
&= \tracedist{\Tr_{\bm{\Ccorr}} \left[\rho^\mathrm{real,virt}_{K_A K_B \CfinalQKD \bm{\Ccorr} \EfinalQKD \wedge \Onice} - \rho^\mathrm{ideal,virt}_{K_A K_B \CfinalQKD \bm{\Ccorr} \EfinalQKD \wedge \Onice}  \right] } \\
&\leq 
\tracedist{\rho^\mathrm{real,virt}_{K_A K_B \CfinalQKD \bm{\Ccorr} \EfinalQKD \wedge \Onice} - \rho^\mathrm{ideal,virt}_{K_A K_B \CfinalQKD \bm{\Ccorr} \EfinalQKD \wedge \Onice} } 
\end{aligned}
\end{equation}
The required claim follows from noting that the above inequality is true for all states $\rho^\mathrm{real}_{K_A K_B \CfinalQKD  \EfinalQKD} \in \worldreal(\coreQKDprotocol)$. In fact, the inequality above can actually be tightened to an equality, since $\bm{\Ccorr}$ is simply a copy of the messages in $\CfinalQKD$ (although the weaker inequality already suffices for our purposes).

\end{proof}

The following lemma reduces the security analysis to states in $\worldvirtual(\coreQKDprotocol)$ (without needing to consider partial states on event $\Onice$).

\begin{lemma} \label{lemma:reductiontwo}
 Consider the same setup as in \cref{theorem:reductionstatement,corr:reduction}. Then, the $\epssecure$-security for all output states in $\worldvirtual(\coreQKDprotocol)$, implies $\epssecure$-security for all output states in $\worldvirtual(\coreQKDprotocol)$, partial on the event $\Onice$. That is 
\begin{equation}
      \begin{aligned}
\tracedist{\rho^\mathrm{real,virt}_{K_A K_B \CfinalQKD \bm{\Ccorr} \EfinalQKD }  - \rho^\mathrm{ideal,virt}_{K_A K_B \CfinalQKD \bm{\Ccorr}  \EfinalQKD } }  &\leq \epssecure \qquad \forall \rho^\mathrm{real,virt}_{K_A K_B \CfinalQKD \bm{\Ccorr} \EfinalQKD } \in  \worldvirtual(\coreQKDprotocol) \\
    &\Downarrow \\
    \tracedist{\rho^\mathrm{real,virt}_{K_A K_B \CfinalQKD \bm{\Ccorr} \EfinalQKD \wedge \Onice }  - \rho^\mathrm{ideal,virt}_{K_A K_B \CfinalQKD \bm{\Ccorr}  \EfinalQKD \wedge \Onice} }  &\leq \epssecure \qquad \forall \rho^\mathrm{real,virt}_{K_A K_B \CfinalQKD \bm{\Ccorr} \EfinalQKD  } \in  \worldvirtual(\coreQKDprotocol) \\
      \end{aligned}
  \end{equation}
\end{lemma}

\begin{proof}
For any state $\rho^\mathrm{real,virt}_{K_A K_B \CfinalQKD \bm{\Ccorr} \EfinalQKD  } \in  \worldvirtual(\coreQKDprotocol) $ we have
\begin{equation}
\begin{aligned}
&\tracedist{\rho^\mathrm{real,virt}_{K_A K_B \CfinalQKD \bm{\Ccorr} \EfinalQKD \wedge \Onice }  - \rho^\mathrm{ideal,virt}_{K_A K_B \CfinalQKD \bm{\Ccorr}  \EfinalQKD  \wedge \Onice } }  \\
&\leq \tracedist{\rho^\mathrm{real,virt}_{K_A K_B \CfinalQKD \bm{\Ccorr} \EfinalQKD \wedge \Onice }  - \rho^\mathrm{ideal,virt}_{K_A K_B \CfinalQKD \bm{\Ccorr}  \EfinalQKD  \wedge \Onice} }  \\
&+\tracedist{\rho^\mathrm{real,virt}_{K_A K_B \CfinalQKD \bm{\Ccorr} \EfinalQKD \wedge \Onice^\complement}  - \rho^\mathrm{ideal,virt}_{K_A K_B \CfinalQKD \bm{\Ccorr}  \EfinalQKD  \wedge \Onice^\complement} } \\
&= \tracedist{\rho^\mathrm{real,virt}_{K_A K_B \CfinalQKD \bm{\Ccorr} \EfinalQKD}  - \rho^\mathrm{ideal,virt}_{K_A K_B \CfinalQKD \bm{\Ccorr}  \EfinalQKD  } }
\end{aligned}
\end{equation}
where the first inequality follows from the fact that we add positive terms to the right hand side of the inequality, and the final equality follows from the fact that the states conditioned on $\Onice$ and $\Onice^\complement$ have support on orthogonal spaces.
\end{proof}

The following lemma reduces the security analysis to states in $\worldhonest(\coreQKDprotocol)$.

\begin{lemma} \label{lemma:reductionthree}
Consider the same setup as in \cref{theorem:reductionstatement,corr:reduction}. Then, the $\epssecure$-security for all output states in $\worldhonest(\coreQKDprotocol)$  implies $\epssecure$-security for all output states in $\worldvirtual(\coreQKDprotocol)$. That is 
\begin{equation}
      \begin{aligned}
\tracedist{\rho^\mathrm{real,hon}_{K_A K_B \CfinalQKD  \EfinalQKD }  - \rho^\mathrm{ideal,hon}_{K_A K_B \CfinalQKD   \EfinalQKD } }  &\leq \epssecure \qquad \forall \rho^\mathrm{real,hon}_{K_A K_B \CfinalQKD  \EfinalQKD } \in  \worldhonest(\coreQKDprotocol) \\
    &\Downarrow  \\
\tracedist{\rho^\mathrm{real,virt}_{K_A K_B \CfinalQKD \bm{\Ccorr} \EfinalQKD  }  - \rho^\mathrm{ideal,virt}_{K_A K_B \CfinalQKD \bm{\Ccorr}  \EfinalQKD } }  &\leq \epssecure \qquad \forall \rho^\mathrm{real,virt}_{K_A K_B \CfinalQKD \bm{\Ccorr} \EfinalQKD  } \in  \worldvirtual(\coreQKDprotocol) \\
      \end{aligned}
  \end{equation}
\end{lemma}
The idea behind the proof is based on the intuition that, in the virtual setting, the registers $\bm{\Ccorr}$ capture the communication that would occur in the honest model, and Alice and Bob use these registers for determining their actions. Consequently, the actual received messages are irrelevant --- in fact, we can even assume that they are never delivered.
\begin{proof}

Consider any state $\rho^\mathrm{real,virt}_{K_A K_B \CfinalQKD  \bm{\Ccorr} \EfinalQKD} \in \worldvirtual(\coreQKDprotocol)$, and fix the corresponding attack strategy by Eve. From our construction of the virtual authentication setting, this implies that messages in $\bm{\Ccorr}$ are always correct and are received some time after being sent. Messages in $\CfinalQKD$ may include $\authabort$s and timing irregularities, but they are not used by Alice or Bob in the protocol at all. 

Next, consider the related attack strategy in the honest authentication setting, which we construct from the attack strategy in the virtual setting. Here Eve creates a copy of every message sent out from Alice’s or Bob’s laboratory. Each message is correctly delivered to the receiving party (according to the timing specified by $\bm{\Ccorr}$). Instead of attacking those messages, Eve performs her attack from the virtual setting on the \emph{copy} of the messages sent out, and does not forward the resulting (potentially tampered) message to the receiving party (see \cref{fig:connectingworlds}). Let the resulting state be $\rho^\mathrm{real,hon}_{K_A K_B \CfinalQKD \EfinalQKD} \in \worldhonest(\coreQKDprotocol)$. Then, we have the following equality with a slight abuse of notation:
\begin{equation}
    \rho^\mathrm{real,virt}_{K_A K_B \CfinalQKD  \bm{\Ccorr} \EfinalQKD} 
    = 
    \rho^\mathrm{real,hon}_{K_A K_B \CfinalQKD \EfinalQKD},
\end{equation}
where $\CfinalQKD$ and $\EfinalQKD$ on the left-hand side (which store the outcome of Eve’s attack) are identified with $\EfinalQKD$ on the right-hand side (where Eve implements the same attack on a copy of the classical messages and does not forward the resulting message to Bob), and $\bm{\Ccorr}$ on the left-hand side (which contains the correctly delivered, correctly timed messages) is identified with $\CfinalQKD$ on the right-hand side (the honest setting, where correct messages are received at the correct times).
\begin{figure}[h!]
  \centering
  \begin{tikzpicture}[every node/.style={inner sep=1pt}]

    \begin{scope}[xshift=-5.5cm]
      \node[participant] (AliceL) {Alice};
      \node[participant] (EveL)   [right=2cm of AliceL] {Eve};
      \node[participant] (BobL)   [right=2cm of EveL]   {Bob};

      \draw[lifeline] (AliceL) -- ++(0,-5);
      \draw[lifeline] (EveL)   -- ++(0,-5);
      \draw[lifeline] (BobL)   -- ++(0,-5);

      \coordinate (A0L) at ($(AliceL)+(0,-\rowsep)$);
      \coordinate (B0L) at ($(BobL)+(0,-\rowsep)$);
       \coordinate (E0L) at ($(EveL)+(0,-\rowsep)$);

      \node[attackbox,align=center,minimum height=4.5cm,anchor=north] (EboxL)
        at ($(EveL)+(0,-0.7)$) {Eve's attack\\in the virtual \\world. \\ \\ $\Ccorr$ contains \\ correct message. \\  };

      \nextrow{A1L}{A0L}
      \nextrow{B1L}{B0L}
      \nextrow{E1L}{E0L}
      \draw[message] (EboxL.east |- A1L) -- node[midway,above] {$C_{E\to B}^{(5)}$} (B1L);

      \nextrow{A2L}{A1L}
      \nextrow{B2L}{B1L}
      \nextrow{E2L}{E1L}

       \draw[message]
        (A2L) -- (B2L);
        
      \draw[message] (A2L) -- node[midway,above] {$C_{A\to E}^{(5)}$} (EboxL.west |- B2L);
      \draw[message] (EboxL.east |- A2L) -- node[midway,above] {$\Ccorr_{E\to B}^{(5)}$} (B2L);
    \end{scope}

    \node[font=\Huge] at (2,-2.5) {$\approx$};

    \begin{scope}[xshift=4cm]
      \node[participant] (AliceR) {Alice};
      \node[participant] (EveR)   [right=2cm of AliceR] {Eve};
      \node[participant] (BobR)   [right=2cm of EveR]   {Bob};

      \draw[lifeline] (AliceR) -- ++(0,-5);
      \draw[lifeline] (EveR)   -- ++(0,-5);
      \draw[lifeline] (BobR)   -- ++(0,-5);

      \coordinate (A0R) at ($(AliceR)+(0,-\rowsep)$);
      \coordinate (B0R) at ($(BobR)+(0,-\rowsep)$);
    \coordinate (E0R) at ($(EveR)+(0,-\rowsep)$);

      \node[attackbox,align=center,minimum height=4.5cm,anchor=north] (EboxR)
        at ($(EveR)+(0,-0.7)$) {Eve's attack \\in the honest\\world \\ where she attacks \\ a copy of \\ sent message \\ \\ };

      \nextrow{A1R}{A0R}
      \nextrow{B1R}{B0R}
      \nextrow{E1R}{E0R}

      \nextrow{A2R}{A1R}
      \nextrow{B2R}{B1R}
      \nextrow{E2R}{E1R}

       \draw[message]
        (A2R) --  (B2R);

\draw[message] 
  (A2R) -- node[midway,above,align=center] {$C_{A\to E}^{(5)}$ \\ + copy} 
  (EboxR.west |- A2R);
      \draw[message] (EboxR.east |- A2R) -- node[midway,above] {$C_{E\to B}^{(5)}$} (B2R);

    \end{scope}

  \end{tikzpicture}
 \caption{For every operation that Eve performs in the virtual authentication setting, one can construct an equivalent operation in the honest authentication setting, where Eve creates a copy of each sent message and performs her original operation on this copy. She does not forward the result of her attack on the copy to the receiver; instead, she forwards the original, unmodified message to the receiving party.}
  \label{fig:connectingworlds}
\end{figure}
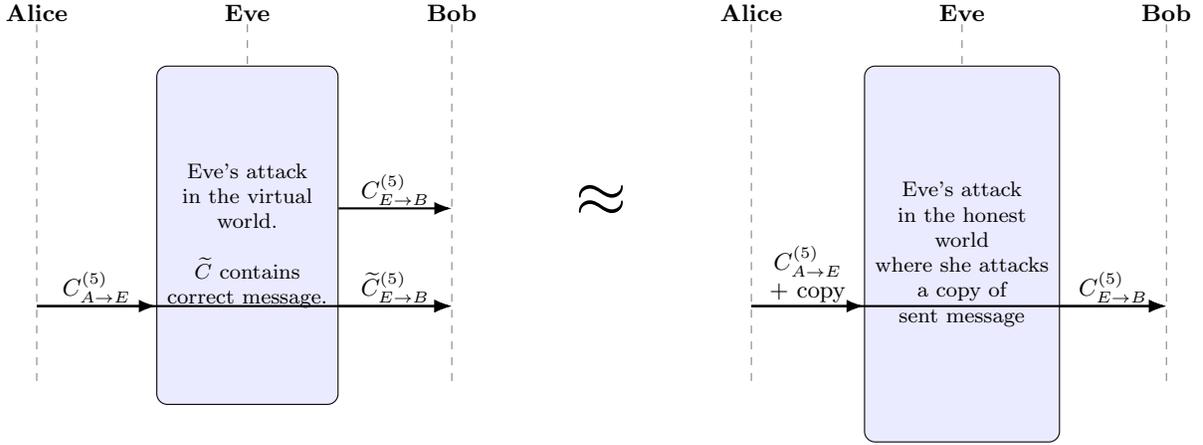

From these properties, we have  
\begin{equation}
\tracedist{
\rho^\mathrm{real,virt}_{K_A K_B \CfinalQKD  \bm{\Ccorr} \EfinalQKD} 
- 
\rho^\mathrm{ideal,virt}_{K_A K_B \CfinalQKD  \bm{\Ccorr} \EfinalQKD}
}
=
\tracedist{
\rho^\mathrm{real,hon}_{K_A K_B \CfinalQKD \EfinalQKD}
-
\rho^\mathrm{ideal,hon}_{K_A K_B \CfinalQKD \EfinalQKD}
}.
\end{equation}

This proves the required statement: if the right-hand side is upper bounded by $\epssecure$ for all states $\rho^\mathrm{real,hon}_{K_A K_B \CfinalQKD \EfinalQKD} \in \worldhonest(\coreQKDprotocol)$, then the same bound holds for the left-hand side for all states $\rho^\mathrm{real,virt}_{K_A K_B \CfinalQKD \bm{\Ccorr} \EfinalQKD} \in \worldvirtual(\coreQKDprotocol)$. This concludes the proof.
\end{proof}

Thus, combining all these reductions, we obtain the required statement which we restate below:

\reductionstatement*
\begin{proof}
    The proof follows from \cref{corr:reduction,lemma:reductionone,lemma:reductiontwo,lemma:reductionthree}.
\end{proof}

\section{Delayed Authentication}
\label{sec:delayedauthentication}

The protocols considered so far in this work utilize an authenticated classical channel for every message sent and received, as described in \cref{subsec:classicalcommunicationsmodel}. However, the number of bits required to authenticate an $n$-bit message with information-theoretic security increases with $n$ (albeit only logarithmically)~\cite{wegman_new_1981,fung_practical_2010}. Consequently, it is more efficient (in terms of authentication key consumption) to batch multiple messages together before authenticating them.

In this section, we consider a modified version of the previously studied scenario, in which the authenticated classical channel is used only twice --- once by each party --- to authenticate the entire transcript of classical communication. For this section, we assume that Alice and Bob have synchronized clocks. We emphasize that this assumption is made only for this section and is necessary to ensure that Alice and Bob can compare transcripts along with their associated timestamps. We again let $\coreQKDprotocol$ denote a generic QKD protocol. However, we now analyze the security of a setting where all communication during $\coreQKDprotocol$ takes place over an insecure, unauthenticated classical channel. In this case, Eve is free to replace any message or arbitrarily modify its timing, and the delivery restrictions described in \cref{subsec:classicalcommunicationsmodel} no longer apply.

We then introduce a delayed Authentication Post-Processing Protocol, denoted by $\delayedAPPprotocol$, described in \cref{subsec:delAPPprotocol}. In this step, Alice and Bob exchange their entire communication transcripts (along with timestamps) over an authenticated channel and verify their consistency. The restrictions described in \cref{subsec:classicalcommunicationsmodel} apply to these messages. If the transcripts match --- i.e., if they confirm that they received the correct messages at the appropriate times --- they proceed; otherwise, they abort the protocol.

With these modifications, we are able to establish an analogue of \cref{theorem:reductionstatement}, stated in \cref{theorem:delayedreductionstatement}. The proof follows via analogous steps to that of \cref{theorem:reductionstatement}. We note that the idea of delayed authentication has been explored in the context of QKD in several prior works \cite{kon_quantumauthenticated_2024,kiktenko_lightweight_2020}. However, these studies typically focus on estimating authentication costs or perform the security analysis only for specific QKD protocols.

For ease of explanation and to maintain intuitive correspondence with earlier sections, we make a slight abuse of notation in this section. In particular, the real output states are denoted using the same symbols as before ($\rho^\mathrm{real}_{K_A K_B \CfinalQKD \EfinalQKD}$), but they now correspond to a different setting $\worlddelreal(\delayedAPPprotocol \circ \coreQKDprotocol)$ (described in greater detail in the following subsections). The honest and virtual settings, however, are defined exactly as before.

\subsection{Delayed Authentication Post-Processing Protocol (del-APP)} \label{subsec:delAPPprotocol}
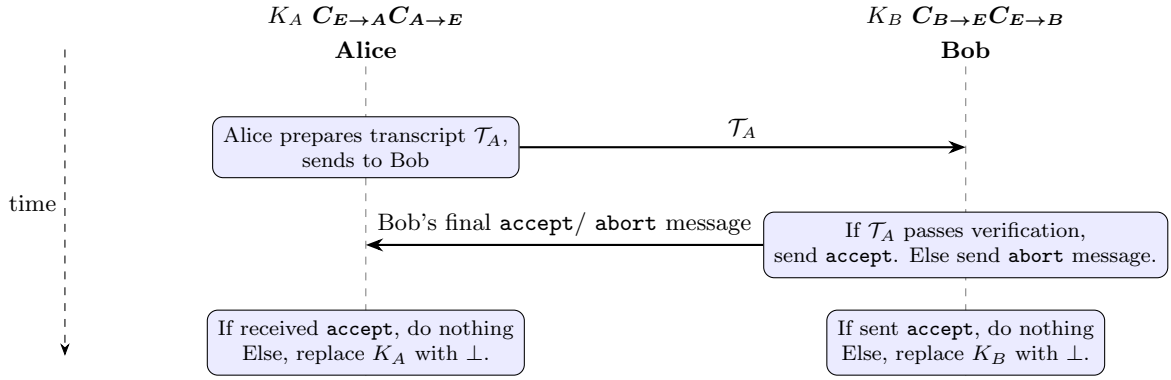
\begin{figure} 
\begin{tikzpicture}[
    >=Stealth,
    participant/.style={font=\small\bfseries},
    lifeline/.style  ={gray,dashed},
    message/.style   ={thick,-Stealth},
    localop/.style   ={draw,rectangle,rounded corners,
                       fill=blue!8,inner sep=4pt,
                       font=\footnotesize,align=center},
]

\def\rowsep{1.3} 

\node[participant] (Alice) {Alice};
\node[participant] (Bob)   [right=8cm of Alice] {Bob};

\node at ($(Alice)+(0,0.45)$) {$K_A\ \CAr \CAs$};
\node at ($(Bob)  +(0,0.45)$) {$K_B\ \CBs \CBr$};

\draw[lifeline] (Alice) -- ++(0,-5.0);
\draw[lifeline] (Bob)   -- ++(0,-5.0);

\coordinate (A0) at ($(Alice)+(0,0)$);
\coordinate (B0) at ($(Bob)  +(0,0)$);

\nextrow{A1}{A0}
\nextrow{B1}{B0}

\node[localop] (A1op) at (A1) {Alice prepares transcript $\transcript_A$, \\ sends to Bob};
\draw[message] (A1op) -- node[midway,above] {$\transcript_A$} (B1);

\nextrow{A2}{A1}
\nextrow{B2}{B1}

\node[localop] (B2op) at (B2) {If $\transcript_A$ passes verification, \\
send \accept. Else send \abort message.};
\draw[message] (B2op) -- node[midway,above] {Bob's final \accept / \abort message} (A2);

\nextrow{A3}{A2}
\nextrow{B3}{B2}

\node[localop] (A3op) at (A3) {If received \accept, do nothing \\
Else, replace $K_A$ with $\bot$.};
\node[localop] (B3op) at (B3) {If sent \accept, do nothing \\
Else, replace $K_B$ with $\bot$.};

\draw[->, dashed]
  ($(Alice)+(-4,0)$) -- ($(AliceEnd)+(-4,-1)$)
  node[midway, left] {time};

\end{tikzpicture}
\caption{Schematic of \nameref{prot:delauthpp}  described in \cref{subsec:delAPPprotocol}. Alice and Bob first communicate and verify transcripts. If transcripts match, they accept the protocol. Else, they abort the protocol and replace their key registers with $\bot$s.}
\label{fig:delappprotocol}
\end{figure}

As before, let $\rho^\mathrm{real}_{K_A K_B \CfinalQKD \EfinalQKD}$ denote the final state obtained after the execution of the core QKD protocol $\coreQKDprotocol$, where $\CfinalQKD = \CAs \CBr \CBs \CAr$ collects all classical communication that occurred during the core protocol, and $\EfinalQKD$ denotes all of Eve’s side information, which may include a copy of the classical communication. Recall our notation: Alice sends (receives) her $i$th message in the register $C^{(i)}_{A \rightarrow E}$ ($C^{(i)}_{E\rightarrow A}$) at time $t^{(i)}_{A \rightarrow E}$ ($t^{(i)}_{E \rightarrow A}$) with Bob's messages and timings defined analogously. During the authentication post-processing phase, Alice and Bob start with the state
$\rho^\mathrm{real}_{K_A K_B \CfinalQKD \EfinalQKD}$ and perform the following actions (see  \cref{fig:delappprotocol}).

\vspace{2em}
\begin{prot}[del-AuthPP Protocol]
\label{prot:delauthpp}
Starts with the state
$\rho^\mathrm{real}_{K_A K_B \CfinalQKD \EfinalQKD}$, after the core QKD protocol has finished.
Classical communication is undertaken in the registers
$\CAsauth, \CBrauth, \CBsauth, \CArauth$.

\begin{enumerate}[label=\textbf{dAPP~\arabic*}, ref=dAPP~\arabic*]
\item \label{prot:delauthpp-alice}
Alice prepares a transcript $\transcript_A$ describing the messages she sent
and received during $\coreQKDprotocol$, and their timings. 
That is, she prepares
\[
\transcript_A =
\left(
\left\{ \left( C^{(i)}_{A \rightarrow E}, t^{(i)}_{A \rightarrow E} \right) \right\}_i,
\left\{ \left( C^{(j)}_{E \rightarrow A}, t^{(j)}_{E \rightarrow A} \right) \right\}_j
\right).
\]
She sends $\transcript_A$ to Bob by making one use of the  authenticated
classical channel.

\item \label{prot:delauthpp-bob}
If Bob receives an $\authabort$, he sends an $\abort$ message to Alice by making one use of the authenticated classical channel.
Otherwise, he checks whether the received transcript matches his own.
That is, he verifies that messages were received after they were sent, and that the message contents match. Formally he checks whether:
\begin{equation} \label{eq:timeandmessagechecks}
  \begin{aligned}
    t^{(i)}_{A \rightarrow E} &\leq t^{(i)}_{E \rightarrow B} 
    &&\forall i, \\
    t^{(i)}_{B \rightarrow E} &\leq t^{(i)}_{E \rightarrow A} 
    &&\forall i, \\
    C^{(i)}_{A \rightarrow E} \text{ and }& C_{E\rightarrow B}^{(i)} \text{ contain identical messages }
    &&\forall i, \\
    C_{E \rightarrow A}^{(i)} \text{ and }&  C_{B \rightarrow E}^{(i)}  \text{ contain identical messages }
    &&\forall i.
  \end{aligned}
\end{equation}
If verification passes, he sends an $\accept$ message;
otherwise, he sends an $\abort$ message by making one use of the authenticated classical channel.

\item \label{prot:delauthpp-bob2}
If Bob sends an $\accept$ message, he does nothing.
If he sends an $\abort$ message, he replaces $K_B$ with $\bot$.
We denote by $l_B$ the final key length stored in $K_B$.

\item \label{prot:delauthpp-alice2}
If Alice receives an $\accept$ message, she does nothing.
If Alice receives either an $\authabort$ or an $\abort$ message,
she replaces her key register $K_A$ with $\bot$.
We denote by $l_A$ the final key length stored in $K_A$.

\end{enumerate}
\end{prot}
\vspace{2em}

%

Mathematically, the above protocols can be described as follows:
\begin{enumerate}
    \item In \cref{prot:delauthpp-alice} and \cref{prot:delauthpp-bob}, Alice and Bob engage in two rounds of authenticated classical communication. This can be described by a map (influenced by Eve) $\delauthcommmap \in \CPTP(\CfinalQKD \EfinalQKD,\Cfinal \Efinal)$.
    \item In \cref{prot:delauthpp-alice2,prot:delauthpp-bob2}, Alice and Bob  use the result of the communication in the previous step $\Cauth$ to determine their final $\accept$ / $\abort$ status. This is described as a map 
    $\delauthupdatemap \in \CPTP( K_A K_B \Cauth, K_A K_B \Cauth )$
\end{enumerate}
The final output state is denoted by  
\begin{equation} \label{eq:delayedtimethingy}
    \rho^\mathrm{real,final}_{K_A K_B \Cfinal \Efinal} = \delauthupdatemap \circ \delauthcommmap \left[\rho^\mathrm{real}_{K_A K_B \CfinalQKD \EfinalQKD}\right],
\end{equation}
 where $\Cfinal = \CfinalQKD \Cauth $.
\begin{remark}
    Since we can use synchronized clocks in this scenario, we can simply declare that the \nameref{prot:delauthpp} starts only after some fixed time marking the end of the core protocol $\coreQKDprotocol$ has passed. Thus, we can directly write \cref{eq:delayedtimethingy} without conditioning on any events, or needing to include extra messages as was required for \cref{eq:APPisaftercore}.
\end{remark}

Notice that the structure of this protocol closely mirrors that of \nameref{prot:authpp}, except that the first two steps of \nameref{prot:authpp} are omitted here. This alignment ensures that the ensuing proofs proceed in essentially the same way.

\begin{remark} \label{remark:memory}
Notice that in this setting, there are only two uses of the authenticated classical channel. Consequently, the amount of secret key required for authentication is significantly lower than what was considered previously in this work. However, Alice and Bob must now store all messages sent and received (along with their corresponding timestamps) until the very end of the protocol, in order to compare and match their transcripts. This, in turn, reduces the benefits of on-the-fly announcements \cite[Remark 3.1]{tupkary2025qkdsecurityproofsdecoystate}, which allow Alice and Bob to perform public announcements on-the-fly and save  classical memory and storage requirements.
Nevertheless, one can still envision a use case where on-the-fly announcements are performed along with delayed authentication to enable Alice and Bob to begin their classical processing (such as sifting) earlier, while still retaining the necessary information in memory for transcript comparison at a later stage. Alternatively, one can accept the requirement for larger classical memory (and not implement on-the-fly announcements), in order to save authentication keys. Finally, note that Alice and Bob need only store sufficient information to recover the ordering of all messages; the precise timestamps are not necessary. This can simplify the storage requirements further.
\end{remark}

\begin{remark} \label{remark:notsendingtranscript}
 One may be tempted to reduce the communication cost of \nameref{prot:delauthpp}. As written, Alice sends Bob her entire transcript, which contains every classical message of the protocol and its timestamp and can therefore be very long. Since an authenticated channel is normally realized by attaching to each message a short tag, it is natural to ask whether Alice could send Bob only the tag of her transcript, with Bob accepting whenever it equals the tag he computes from his own transcript. This looks attractive because the tag length is much shorter than the transcript length.
 
The catch is that such a test accepts only when Alice's and Bob's transcripts are identical, so it can work only if the two transcripts coincide in every honest run. If the transcript records the time of each message, they will never agree (unless some mechanism ensures the time delay in message transmission is much smaller than the precision of the synchronized clocks). One might then try to strip out the timestamps and keep only the order of the messages, recording each message by its position in the sequence each party sends and receives. In this case, the two honest transcripts, and hence their tags, become identical. However, in this case, there is no check to ensure that the timing conditions from \cref{eq:timeandmessagechecks} are satisfied. In fact, we can explicitly construct a scenario in  Appendix~\ref{app:timing-attack} where the ordering of messages from the point of view of each party remains unchanged, but Eve still manages to completely recover the secret key, violating the conditions from \cref{eq:timeandmessagechecks}. 

One possibility may be to have each party include the sending time of the message in the message itself, and then simply send a tag of the transcript of all sent messages (which includes the timing) at the end of the protocol. This allows the receiving party to verify the authenticity of the received timestamps, and check that they received messages after they were sent. Other variations, under different assumptions, may also be applicable. We leave such considerations to future work.
\end{remark} 

\subsection{Reduction Statement} \label{subsec:delayedreductionstatement}
We are interested in the security analysis of $\delayedQKDprotocol = \delayedAPPprotocol \circ \coreQKDprotocol$.
Let us denote the set of output states that are possible when we use insecure, unauthenticated communication during $\coreQKDprotocol$ and authenticated communication during $\delayedAPPprotocol$ as $\worlddelreal(\delayedQKDprotocol)$.  The analogous result  to \cref{theorem:reductionstatement} for the delayed authentication setting can now be stated and proved  in an analogous manner.

\begin{restatable}[Reduction of QKD security analysis to the honest authentication setting with delayed authentication]{theorem}{delayedreductionstatement}
\label{theorem:delayedreductionstatement}
Let $\coreQKDprotocol$ be an arbitrary QKD protocol. Let $\delayedAPPprotocol$ be the \nameref{prot:delauthpp} described in \cref{subsec:delAPPprotocol}, executed after the core QKD protocol $\coreQKDprotocol$. Let $\delayedQKDprotocol = \delayedAPPprotocol \circ \coreQKDprotocol$ denote the resulting QKD protocol. Let $\worlddelreal(\delayedQKDprotocol)$ denote the set of output states in the delayed authentication setting, where communication during $\coreQKDprotocol$ is not authenticated while communication during $\delayedAPPprotocol$ is authenticated (see \cref{subsec:classicalcommunicationsmodel}). Let $\worldhonest(\coreQKDprotocol)$ denote the set of output states in the honest authentication setting (see \cref{subsec:classicalcommunicationsmodel}).
Then, the $\epssecure$-security for all output states in $\worldhonest(\coreQKDprotocol)$ implies $\epssecure$-security for all output states in $\worlddelreal(\delayedQKDprotocol)$. That is,
\begin{equation}
    \begin{aligned}
          \tracedist{ \rho^\mathrm{real,hon}_{K_A K_B \CfinalQKD \EfinalQKD} -  \rho^\mathrm{ideal,hon}_{K_A K_B \CfinalQKD \EfinalQKD}} &\leq \epssecure \qquad \forall \rho^\mathrm{real}_{K_A K_B \CfinalQKD \EfinalQKD} \in \worldhonest(\coreQKDprotocol) \\
          &\Downarrow \\
     \tracedist{ \rho^\mathrm{real,final}_{K_A K_B \Cfinal \Efinal} -  \rho^\mathrm{ideal,final}_{K_A K_B \Cfinal \Efinal}} &\leq \epssecure \qquad \forall \rho^\mathrm{real,final}_{K_A K_B \Cfinal \Efinal} \in \worlddelreal(\delayedQKDprotocol).
    \end{aligned}
\end{equation}
\end{restatable}

The proof is stated in Appendix~\ref{appendix:delayedreductionproof}.

\section{Conclusion} \label{sec:conclusion}
In summary, we addressed the problem of analyzing QKD protocols in the {\realistic} authentication setting, where authentication may result in asymmetric aborts and message timing can be modified. This setting necessitates a modification of the standard QKD security definition and, crucially, renders nearly all existing analyses, which are undertaken in the honest authentication setting, invalid under {\realistic} authentication assumptions.

By introducing an authentication post-processing protocol, we demonstrated that the security of a complete QKD protocol (including the authentication post-processing) in the {\realistic} authentication setting can be reduced to that of the core QKD protocol in the honest authentication setting. The latter is the setting overwhelmingly considered in existing QKD security proofs. Hence, our results retroactively lift the security of such prior analyses to the {\realistic} authentication setting.  As a concrete example, our result can be applied to lift the recently obtained rigorous decoy-state BB84 security proof of Ref.~\cite{mizutani2025protocolleveldescriptionselfcontainedsecurity} to the practical authentication setting, and it has already been incorporated into Ref.~\cite{tupkary2026rigorouscompletesecurityproof}. Moreover, our construction is fully general; it applies to any QKD protocol and is equally suitable for device-dependent, measurement-device-independent\footnote{Such protocols also involve public announcements made by an untrusted third party. We emphasize that our analysis applies only to the communication between Alice and Bob themselves; it does not address how to handle announcements originating from an external third party. Typically in MDI protocols, Alice and Bob need to achieve consensus on the third party’s announcements --- while our work does not directly address those announcements, it does show that Alice and Bob can rely on the honest-authentication scenario for communication between themselves, in order to achieve such consensus.} (MDI), and device-independent (DI) protocols, whether of the prepare-and-measure or entanglement-based type. We also presented two variants of this construction: one in which all communication is authenticated, and another where only the final two messages are authenticated, thereby reducing the consumption of authentication keys.

In this work, our perspective was that Eve’s attack together with the protocol description jointly determine a single channel mapping input states to output states.  Looking ahead, an important next step is to formalize the new security definition and the {\realistic} authenticated channel within a composable security framework. Achieving this will likely require employing models such as causal boxes \cite{portmann_causal_2017} or related formalisms that explicitly capture the causal and temporal structure of classical communication.  

\section*{Acknowledgments}

This article was written as part of the Qu-Gov project, commissioned by the German Federal Ministry of Finance. We thank the Bundesdruckerei - Innovation Leadership and Team for their support and encouragement. We thank Holger Eble of the  Bundesdruckerei and Tobias Hemmert of the Federal Office for Information Security (BSI) for providing valuable feedback and comments on multiple drafts of this manuscript. This work was conducted at the Institute for Quantum Computing, University of Waterloo, which
is funded by the Government of Canada through ISED. DT was partially funded by the Mike and Ophelia Lazaridis Fellowship. 

We are grateful to Norbert L\"utkenhaus for bringing this problem to our attention, and for various discussions on this topic. We thank Renato Renner and Martin Sandfuchs for valuable feedback on the proof strategy employed in this work, and we also thank Martin for useful feedback on the manuscript itself. Finally, we thank Jerome Wiesemann, Aodh\'an Corrigan and Zhiyao Wang for their helpful comments on early drafts of this manuscript. 
We thank the anonymous reviewer for pointing out a subtle flaw due to implicit timing assumptions in our original proof, which allowed us to provide the fix described in \cref{remark:abortduetodelay}. We thank Federico Grasselli for some interesting discussions that led to \cref{remark:notsendingtranscript}.

\bibliographystyle{IEEEtran}
\bibliography{bibliography}

@misc{tupkary2026rigorouscompletesecurityproof,
      title={A rigorous and complete security proof of decoy-state BB84 quantum key distribution}, 
      author={Devashish Tupkary and Shlok Nahar and Amir Arqand and Ernest Y. -Z. Tan and Norbert Lütkenhaus},
      year={2026},
      eprint={2601.18035},
      archivePrefix={arXiv},
      primaryClass={quant-ph},
      url={https://arxiv.org/abs/2601.18035}, 
}

@misc{tupkary2025qkdsecurityproofsdecoystate,
      title={{Security proofs for practical QKD: variations, techniques, gaps, and limitations}}, 
      author={Devashish Tupkary and Ernest Y. -Z. Tan and Shlok Nahar and Lars Kamin and Norbert Lütkenhaus},
      year={2026},
      eprint={2502.10340},
      archivePrefix={arXiv},
      primaryClass={quant-ph},
      url={https://arxiv.org/abs/2502.10340}, 
}

@misc{ferradini2025definingsecurityquantumkey,
      title={Defining Security in Quantum Key Distribution}, 
      author={Carla Ferradini and Martin Sandfuchs and Ramona Wolf and Renato Renner},
      year={2025},
      eprint={2509.13405},
      archivePrefix={arXiv},
      primaryClass={quant-ph},
      url={https://arxiv.org/abs/2509.13405}, 
}

@article{fung_practical_2010,
  title = {Practical Issues in Quantum-Key-Distribution Postprocessing},
  author = {Fung, Chi-Hang Fred and Ma, Xiongfeng and Chau, H. F.},
  year = {2010},
  month = jan,
  journal = {Physical Review A},
  volume = {81},
  number = {1},
  pages = {012318},
  issn = {1050-2947, 1094-1622},
  doi = {10.1103/PhysRevA.81.012318},
  urldate = {2024-01-17},
  langid = {english},
  language = {en}
}

@article{kiktenko_lightweight_2020,
  title = {Lightweight {{Authentication}} for {{Quantum Key Distribution}}},
  author = {Kiktenko, Evgeniy O. and Malyshev, Aleksei O. and Gavreev, Maxim A. and Bozhedarov, Anton A. and Pozhar, Nikolay O. and Anufriev, Maxim N. and Fedorov, Aleksey K.},
  year = {2020},
  month = oct,
  journal = {IEEE Transactions on Information Theory},
  volume = {66},
  number = {10},
  pages = {6354--6368},
  issn = {1557-9654},
  doi = {10.1109/TIT.2020.2989459},
  urldate = {2024-12-17}
}

@article{broadbent_2023,
   title={Categorical composable cryptography: extended version},
   volume={Volume 19, Issue 4},
   ISSN={1860-5974},
   url={http://dx.doi.org/10.46298/lmcs-19(4:30)2023},
   DOI={10.46298/lmcs-19(4:30)2023},
   journal={Logical Methods in Computer Science},
   publisher={Centre pour la Communication Scientifique Directe (CCSD)},
   author={Broadbent, Anne and Karvonen, Martti},
   year={2023},
   month=dec }

@inproceedings{maurer_abstract_2011,
  title = {Abstract {{Cryptography}}},
  booktitle = {International {{Conference}} on {{Supercomputing}}},
  author = {Maurer, U. and Renner, R.},
  year = {2011},
  urldate = {2024-12-18}
}

@misc{mizutani2025protocolleveldescriptionselfcontainedsecurity,
      title={Protocol-level description and self-contained security proof of decoy-state BB84 QKD protocol}, 
      author={Akihiro Mizutani and Toshihiko Sasaki and Go Kato},
      year={2025},
      eprint={2504.20417},
      archivePrefix={arXiv},
      primaryClass={quant-ph},
      url={https://arxiv.org/abs/2504.20417}, 
}

@article{portmann_causal_2017,
  title = {Causal {{Boxes}}: {{Quantum Information-Processing Systems Closed Under Composition}}},
  shorttitle = {Causal {{Boxes}}},
  author = {Portmann, Christopher and Matt, Christian and Maurer, Ueli and Renner, Renato and Tackmann, Bj{\"o}rn},
  year = {2017},
  month = may,
  journal = {IEEE Transactions on Information Theory},
  volume = {63},
  number = {5},
  pages = {3277--3305},
  issn = {1557-9654},
  doi = {10.1109/TIT.2017.2676805},
  urldate = {2024-12-18}
}

@article{portmann_key_2014,
  title = {Key {{Recycling}} in {{Authentication}}},
  author = {Portmann, Christopher},
  year = {2014},
  month = jul,
  journal = {IEEE Transactions on Information Theory},
  volume = {60},
  number = {7},
  pages = {4383--4396},
  issn = {1557-9654},
  doi = {10.1109/TIT.2014.2317312},
  urldate = {2024-12-02}
}

@article{portmann_security_2022,
  title = {Security in Quantum Cryptography},
  author = {Portmann, Christopher and Renner, Renato},
  year = {2022},
  month = jun,
  journal = {Reviews of Modern Physics},
  volume = {94},
  number = {2},
  pages = {025008},
  publisher = {American Physical Society},
  doi = {10.1103/RevModPhys.94.025008},
  urldate = {2024-12-02}
}

@phdthesis{renner_security_2005,
  type = {Doctoral {{Thesis}}},
  title = {Security of Quantum Key Distribution},
  author = {Renner, Renato},
  year = {2005},
  doi = {10.3929/ethz-a-005115027},
  urldate = {2024-10-05},
  copyright = {http://rightsstatements.org/page/InC-NC/1.0/},
  langid = {english},
  language = {en},
  school = {ETH Zurich}
}

@article{tomamichel_largely_2017,
  title = {{A Largely Self-Contained and Complete Security Proof for Quantum Key Distribution}},
  author = {Tomamichel, Marco and Leverrier, Anthony},
  year = {2017},
  month = jul,
  journal = {Quantum},
  volume = {1},
  pages = {14},
  publisher = {Verein zur F{\"o}rderung des Open Access Publizierens in den Quantenwissenschaften},
  doi = {10.22331/q-2017-07-14-14},
  urldate = {2024-10-05},
  langid = {british},
  language = {en-GB}
}

@article{wegman_new_1981,
  title = {New Hash Functions and Their Use in Authentication and Set Equality},
  author = {Wegman, Mark N. and Carter, J. Lawrence},
  year = {1981},
  month = jun,
  journal = {Journal of Computer and System Sciences},
  volume = {22},
  number = {3},
  pages = {265--279},
  issn = {0022-0000},
  doi = {10.1016/0022-0000(81)90033-7},
  urldate = {2024-12-02}
}

@inproceedings{krawczyk-LFSR-1994,
  title={{LFSR-based Hashing and Authentication}},
  booktitle={Advances in Cryptology - CRYPTO '94, 14th Annual International Cryptology Conference, Santa Barbara, California, USA, August 21-25, 1994, Proceedings},
  series={Lecture Notes in Computer Science},
  publisher={Springer},
  volume={839},
  pages={129-139},
  doi={10.1007/3-540-48658-5_15},
  author={Hugo Krawczyk},
  year=1994
}

@misc{kon_quantumauthenticated_2024,
      title={Quantum Authenticated Key Expansion with Key Recycling}, 
      author={Wen Yu Kon and Jefferson Chu and Kevin Han Yong Loh and Obada Alia and Omar Amer and Marco Pistoia and Kaushik Chakraborty and Charles Lim},
      year={2024},
      eprint={2409.16540},
      archivePrefix={arXiv},
      primaryClass={quant-ph},
      url={https://arxiv.org/abs/2409.16540}, 
}

@inproceedings{maurer_constructive_2012,
    address = {Berlin, Heidelberg},
    title = {Constructive {Cryptography} – {A} {New} {Paradigm} for {Security} {Definitions} and {Proofs}},
    isbn = {978-3-642-27375-9},
    doi = {10.1007/978-3-642-27375-9_3},
    language = {en},
    booktitle = {Theory of {Security} and {Applications}},
    publisher = {Springer},
    author = {Maurer, Ueli},
    editor = {Mödersheim, Sebastian and Palamidessi, Catuscia},
    year = {2012},
    pages = {33--56},
}

@misc{maurer_indifferentiability_2016,
    title = {From {Indifferentiability} to {Constructive} {Cryptography} (and {Back})},
    url = {https://eprint.iacr.org/2016/903},
    urldate = {2025-04-01},
    author = {Maurer, Ueli and Renner, Renato},
    year = {2016},
    note = {Publication info: Published by the IACR in TCC 2016},
}

@InProceedings{ben-or_universal_2004,
author="Ben-Or, Michael
and Horodecki, Micha{\l}
and Leung, Debbie W.
and Mayers, Dominic
and Oppenheim, Jonathan",
editor="Kilian, Joe",
title="The Universal Composable Security of Quantum Key Distribution",
booktitle="Theory of Cryptography",
year="2005",
publisher="Springer Berlin Heidelberg",
address="Berlin, Heidelberg",
pages="386--406",
isbn="978-3-540-30576-7"
}

\appendices

\section{Proof of Reduction \cref{theorem:delayedreductionstatement}} \label{appendix:delayedreductionproof}

To prove this result, we begin by defining the event $\Onicedel$ as the event in which Alice’s transcript $\transcript_A$ is such that it satisfies the verification conditions.
(Note that this refers to the actual transcript itself, not to whether Bob successfully receives it). Thus, $\Onicedel$ is already determined by the time $\coreQKDprotocol$ concludes.  
Intuitively, conditioned on the event $\Onicedel$, we may assume that authentication behaved honestly during the execution of $\coreQKDprotocol$. Analogous to \cref{lemma:bothabort}, we prove the following lemma, which states that if $\Onicedel$ does not occur, then both parties abort.

\begin{lemma} \label{lemma:delbothabort}
Let $\Onicedel$ be the event that Alice’s transcript $\transcript_A$ is such that it satisfies the verification conditions, and let $\rho^\mathrm{real,final}_{K_A K_B \Cfinal \Efinal}$ denote the final output state at the end of the full QKD protocol $\delayedQKDprotocol$. Then the following equality holds
\begin{equation} \label{eq:delniceeventequality}
\rho^\mathrm{real,final}_{K_A K_B \Cfinal \Efinal | \Onicedel^\complement}= \idealmap \left[ \rho^\mathrm{real,final}_{K_A K_B \Cfinal \Efinal | \Onicedel^\complement} \right].
\end{equation}
Therefore, 
\begin{equation} \label{eq:delniceeventinequality}
\tracedist{\rho^\mathrm{real,final}_{K_A K_B \Cfinal \Efinal } - \rho^\mathrm{ideal,final}_{K_A K_B \Cfinal \Efinal }} =
    \tracedist{\rho^\mathrm{real,final}_{K_A K_B \Cfinal \Efinal \wedge \Onicedel} - \rho^\mathrm{ideal,final}_{K_A K_B \Cfinal \Efinal \wedge \Onicedel}} 
\end{equation}
\end{lemma}
\begin{proof}
    The proof follows in a similar manner as the proof of \cref{lemma:bothabort}, by noting that \nameref{prot:delauthpp} ensures that both parties abort when $\Onicedel^\complement$ occurs. 
\end{proof}

Analogous to \cref{lemma:commutationidealauth}, we now show that conditioned on the event $\Onicedel$, the final real and ideal output states (at the end of $\delayedQKDprotocol$) can be obtained by the action of $\delauthupdatemap \circ \delauthcommmap$ on the real and ideal states at the end of the core QKD protocol $\coreQKDprotocol$.

\begin{lemma}[Commutation of $\idealmap$ and \nameref{prot:delauthpp}] \label{lemma:delcommutationidealauth}
Let $\Onicedel$ be the event in which Alice’s transcript $\transcript_A$ is such that it satisfies the verification conditions. Let $\rho^\mathrm{real}_{K_A K_B \CfinalQKD \EfinalQKD | \Onicedel}$ denote the real state at the end of the core QKD protocol  conditioned on $\Onicedel$. Let the following states denote its evolution through  \nameref{prot:delauthpp}: 
\begin{equation}
\begin{aligned}
    \rho^\mathrm{real,comm}_{K_A K_B \CfinalQKD \Cauth  \Efinal| \Onicedel} &\coloneq \delauthcommmap \left[ \rho^\mathrm{real}_{K_A K_B \CfinalQKD \EfinalQKD| \Onicedel} \right], \\
     \rho^\mathrm{real,final}_{K_A K_B \Cfinal\Efinal| \Onicedel} &\coloneq \delauthupdatemap \left[ \rho^\mathrm{real,comm}_{K_A K_B \CfinalQKD \Cauth  \Efinal| \Onicedel} \right].
   \end{aligned}
\end{equation}
Define the corresponding ideal states by the action of the map $\idealmap$ on the real states, i.e 
\begin{equation}
\begin{aligned}
   \rho^\mathrm{ideal}_{K_A K_B \CfinalQKD \EfinalQKD| \Onicedel} &\coloneq \idealmap \left[   \rho^\mathrm{real}_{K_A K_B \CfinalQKD \EfinalQKD| \Onicedel} \right], \\
   \rho^\mathrm{ideal,comm}_{K_A K_B \CfinalQKD \Cauth \Efinal| \Onicedel} &\coloneq \idealmap \left[   \rho^\mathrm{real,comm}_{K_A K_B \CfinalQKD \Cauth\Efinal| \Onicedel} \right], \\
     \rho^\mathrm{ideal,final}_{K_A K_B \Cfinal\Efinal| \Onicedel} &\coloneq \idealmap \left[  \rho^\mathrm{real,final}_{K_A K_B \Cfinal\Efinal| \Onicedel}  \right].
   \end{aligned}
\end{equation}
Then, the ideal states defined above are the same as those obtained by evolving  $ \rho^\mathrm{ideal}_{K_A K_B \CfinalQKD \EfinalQKD| \Onicedel}  $ through  the \nameref{prot:delauthpp}, i.e.,
\begin{equation}
\begin{aligned}
    \rho^\mathrm{ideal,comm}_{K_A K_B \CfinalQKD \Cauth  \Efinal| \Onicedel} &= \delauthcommmap \left[ \rho^\mathrm{ideal}_{K_A K_B \CfinalQKD \EfinalQKD| \Onicedel} \right], \\
     \rho^\mathrm{ideal,final}_{K_A K_B \Cfinal\Efinal| \Onicedel} &= \delauthupdatemap \left[\rho^\mathrm{ideal,comm}_{K_A K_B \CfinalQKD \Cauth  \Efinal| \Onicedel} \right].
     \end{aligned}
\end{equation} 
\end{lemma}
\begin{proof}
The fact that
\begin{equation}
    \delauthcommmap \circ \idealmap \left[ \rho^\mathrm{real}_{K_A K_B \CfinalQKD   \EfinalQKD| \Onicedel} \right] = \idealmap \circ \delauthcommmap \left[ \rho^\mathrm{real}_{K_A K_B \CfinalQKD \EfinalQKD  | \Onicedel} \right]
\end{equation}
follows from the fact that $\delauthcommmap \in \CPTP(\CfinalQKD \EfinalQKD, \Cfinal \Efinal)$ and $\idealmap \in \CPTP(K_A K_B , K_A K_B)$: they act on disjoint registers. (This is analogous to the commutation corresponding to the second arrow in the proof of \cref{lemma:commutationidealauth}).

The fact that
\begin{equation}
    \delauthupdatemap \circ \idealmap \left[ \rho^\mathrm{real,comm}_{K_A K_B \CfinalQKD \Cauth  \Efinal| \Onicedel} \right] = \idealmap \circ \delauthupdatemap \left[ \rho^\mathrm{real,comm}_{K_A K_B \CfinalQKD \Cauth  \Efinal| \Onicedel} \right]
\end{equation}
follows from analogous arguments to the commutation corresponding to the third arrow in the proof of \cref{lemma:commutationidealauth}. In particular, we consider each possible combination of final accept / abort decisions, and find that the required commutation holds for all of them.
\end{proof}

Using these results, we obtain the following corollary which is analogous to \cref{corr:reduction}.

\begin{corollary} \label{corr:delayedreduction}
Consider the same setting as in \cref{theorem:delayedreductionstatement}. The $\epssecure$-security for all output states in $\worlddelreal(\coreQKDprotocol)$ partial on $\Onicedel$, implies $\epssecure$-security for all output states in $\worlddelreal(\delayedQKDprotocol)$. That is 
  \begin{equation}
      \begin{aligned}
    \tracedist{\rho^\mathrm{real}_{K_A K_B \CfinalQKD \EfinalQKD \wedge \Onicedel }  - \rho^\mathrm{ideal}_{K_A K_B \CfinalQKD \EfinalQKD \wedge \Onicedel} }  &\leq \epssecure \qquad \forall \rho^\mathrm{real}_{K_A K_B \CfinalQKD \EfinalQKD } \in  \worlddelreal(\coreQKDprotocol) \\
    &\Downarrow \\
     \tracedist{\rho^\mathrm{real,final}_{K_A K_B \Cfinal \Efinal }  - \rho^\mathrm{ideal,final}_{K_A K_B \Cfinal \Efinal } }  &\leq \epssecure \qquad \forall \rho^\mathrm{real,final}_{K_A K_B \Cfinal \Efinal } \in  \worlddelreal(\delayedQKDprotocol) 
      \end{aligned}
  \end{equation}
\end{corollary}
\begin{proof}
    The claim follows from \cref{lemma:delbothabort,lemma:delcommutationidealauth} using arguments identical to those used in the proof of \cref{corr:reduction}.
\end{proof}

We then construct the virtual authentication setting in exactly the same manner as before (see \cref{subsec:virtualsetting}). In the virtual setting, Eve can do any operation she wants on the classical messages, and the correct messages are delivered at correct timings in the special registers $\bm{\Ccorr}$. Moreover, Alice and Bob use these special registers for further steps in the QKD protocol. We then obtain the following lemma. (Note that \textit{technically}, this virtual setting differs slightly from the one considered in \cref{subsec:virtualsetting}, as the classical communication here is unauthenticated. However, this technical distinction does not affect the proof steps or underlying logic.)

\begin{lemma} \label{lemma:delayedreductionone}
        Consider the same setup as in \cref{theorem:delayedreductionstatement,corr:delayedreduction}. Then, the $\epssecure$-security for all output states in $\worldvirtual(\coreQKDprotocol)$ partial on $\Onicedel$, implies $\epssecure$-security for all output states in $\worlddelreal(\coreQKDprotocol)$ partial on $\Onicedel$. That is 
  \begin{equation}
      \begin{aligned}
    \tracedist{\rho^\mathrm{real,virt}_{K_A K_B \CfinalQKD \bm{\Ccorr} \EfinalQKD \wedge \Onicedel }  - \rho^\mathrm{ideal,virt}_{K_A K_B \CfinalQKD \bm{\Ccorr}  \EfinalQKD \wedge \Onicedel} }  &\leq \epssecure \qquad \forall \rho^\mathrm{real,virt}_{K_A K_B \CfinalQKD \bm{\Ccorr} \EfinalQKD } \in  \worldvirtual(\coreQKDprotocol) \\
    &\Downarrow \\
     \tracedist{\rho^\mathrm{real}_{K_A K_B \CfinalQKD \EfinalQKD \wedge \Onicedel}  - \rho^\mathrm{ideal}_{K_A K_B \CfinalQKD \EfinalQKD \wedge \Onicedel} }  &\leq \epssecure \qquad \forall \rho^\mathrm{real}_{K_A K_B \CfinalQKD \EfinalQKD  } \in  \worlddelreal(\coreQKDprotocol) 
      \end{aligned}
  \end{equation}
\end{lemma}
\begin{proof}
  The claim follows by arguments identical to those used in the proof of \cref{lemma:reductionone}.
\end{proof}

The following lemma reduces the security analysis to states in $\worldvirtual(\coreQKDprotocol)$.

\begin{lemma} \label{lemma:delayedreductiontwo}
 Consider the same setup as in \cref{theorem:delayedreductionstatement,corr:delayedreduction}. Then, the $\epssecure$-security for all output states in $\worldvirtual(\coreQKDprotocol)$, implies $\epssecure$-security for all output states in $\worldvirtual(\coreQKDprotocol)$ partial on $\Onicedel$. That is 
\begin{equation}
      \begin{aligned}
\tracedist{\rho^\mathrm{real,virt}_{K_A K_B \CfinalQKD \bm{\Ccorr} \EfinalQKD }  - \rho^\mathrm{ideal,virt}_{K_A K_B \CfinalQKD \bm{\Ccorr}  \EfinalQKD } }  &\leq \epssecure \qquad \forall \rho^\mathrm{real,virt}_{K_A K_B \CfinalQKD \bm{\Ccorr} \EfinalQKD } \in  \worldvirtual(\coreQKDprotocol) \\
    &\Downarrow \\
    \tracedist{\rho^\mathrm{real,virt}_{K_A K_B \CfinalQKD \bm{\Ccorr} \EfinalQKD \wedge \Onicedel }  - \rho^\mathrm{ideal,virt}_{K_A K_B \CfinalQKD \bm{\Ccorr}  \EfinalQKD \wedge \Onicedel} }  &\leq \epssecure \qquad \forall \rho^\mathrm{real,virt}_{K_A K_B \CfinalQKD \bm{\Ccorr} \EfinalQKD  } \in  \worldvirtual(\coreQKDprotocol) \\
      \end{aligned}
  \end{equation}
\end{lemma}
\begin{proof}
    The claim follows by arguments identical to those used in the proof of \cref{lemma:reductiontwo}.
\end{proof}

The following lemma reduces the security analysis to states in $\worldhonest(\coreQKDprotocol)$.

\begin{lemma} \label{lemma:delayedreductionthree}
Consider the same setup as in \cref{theorem:delayedreductionstatement,corr:delayedreduction}. Then, the $\epssecure$-security for all output states in $\worldhonest(\coreQKDprotocol)$  implies $\epssecure$-security for all output states in $\worldvirtual(\coreQKDprotocol)$. That is 
\begin{equation}
      \begin{aligned}
\tracedist{\rho^\mathrm{real,hon}_{K_A K_B \CfinalQKD  \EfinalQKD }  - \rho^\mathrm{ideal,hon}_{K_A K_B \CfinalQKD   \EfinalQKD } }  &\leq \epssecure \qquad \forall \rho^\mathrm{real,hon}_{K_A K_B \CfinalQKD  \EfinalQKD } \in  \worldhonest(\coreQKDprotocol) \\
    &\Downarrow  \\
\tracedist{\rho^\mathrm{real,virt}_{K_A K_B \CfinalQKD \bm{\Ccorr} \EfinalQKD  }  - \rho^\mathrm{ideal,virt}_{K_A K_B \CfinalQKD \bm{\Ccorr}  \EfinalQKD } }  &\leq \epssecure \qquad \forall \rho^\mathrm{real,virt}_{K_A K_B \CfinalQKD \bm{\Ccorr} \EfinalQKD  } \in  \worldvirtual(\coreQKDprotocol) \\
      \end{aligned}
  \end{equation}
\end{lemma}
\begin{proof}
    The claim follows by arguments identical to those used in the proof of \cref{lemma:reductionthree}.
\end{proof}

Bringing it all together, we obtain the proof of the required reduction statement.

\delayedreductionstatement*
\begin{proof}
 The proof follows from \cref{corr:delayedreduction,lemma:delayedreductionone,lemma:delayedreductiontwo,lemma:delayedreductionthree}.
\end{proof}

\section{A message timing-based attack on QKD}
\label{app:timing-attack}

To illustrate that the timing guarantees of classical messages are important to consider for QKD protocols, we describe a concrete attack that
succeeds whenever the temporal ordering of announcements is not enforced. Consider the following (entirely reasonable) announcement structure for a QKD protocol, carried out
after the quantum signals have been transmitted and measured:
\begin{enumerate}
  \item Bob announces which rounds he detected and which he did not; the
        non-detected rounds are discarded, and only the detected rounds are
        retained.
  \item Alice announces her basis choices on the detected rounds; these rounds
        are then partitioned into test rounds and key-generation rounds. 
  \item Bob announces his bases, together with the outcomes on the test rounds
        (used to determine the length of the final output key to be produced) and his bases on the key-generation rounds
        for sifting.
\end{enumerate}
Announcing the detection pattern first is natural for efficiency: it lets Alice
immediately discard the (typically large) fraction of undetected rounds.  The implicit assumption here  is that Alice
reveals her bases only \emph{after} the signals have reached Bob (since Bob has made his announcement) and the signals are outside of Eve's control.

Suppose now that temporal ordering is not enforced, and a party cannot detect that an incoming message was delivered before the
corresponding message was legitimately sent. Consider the scenario where Eve intercepts
all of Alice's signals and stores them in a quantum memory, forwarding
nothing to Bob for the time being. Eve then fabricates a
``detection announcement'' and delivers it to Alice as though it originated from
Bob, claiming detections on a set of rounds of the adversary's choosing. Unable
to recognise this message as premature, Alice treats it as step~1 and responds,
in step~2, by announcing her bases choices. At this point Eve can measure each stored signal in its correct basis and 
recover the encoded bit. Furthermore, she can then forward exactly those states to Bob for which she claimed a detection outcome, and sends vacuum for the remaining rounds. Bob then subsequently also obtains detection outcomes on the same set of rounds that Eve chose in her fabricated message\footnote{This requires the assumption of ideal efficiency detectors.}, and makes his actual announcement on the detection outcomes. Thus, Bob's message is \textit{identical} in content to the fabricated message that Eve first sends to Alice; the only difference is in the time it was generated. In this scenario, Eve obtains full information about the retained rounds while introducing no detectable disturbance, and security is completely broken.

In the context of our work, this attack is forbidden precisely by the requirements we have imposed on our communication
model (\cref{subsec:classicalcommunicationsmodel}): a message received before it was sent results in an \texttt{auth-abort}. In the case of the delayed authentication scenario from \cref{sec:delayedauthentication}, this attack reflects the need to include timestamps for the transcript comparison, and therefore the need for Alice to send her entire transcript (see \cref{remark:notsendingtranscript}).

\end{document}